\documentclass[numberedappendix]{emulateapj}
\usepackage{graphicx}
\newcommand{\epp}{\ensuremath{e^{\pm}}}

\newcommand{\sigmav}{\langle\sigma v\rangle}

\newcommand{\beq}{\begin{equation}}
\newcommand{\eeq}{\end{equation}}
\newcommand{\be}{\begin{eqnarray}}
\newcommand{\ee}{\end{eqnarray}}

\newcommand{\reffig}[1]{Figure \ref{fig:#1}}
\newcommand{\refsec}[1]{\S \ref{sec:#1}}
\newcommand\bp{\begin{figure}[!ht]}
\newcommand\ep{\end{figure}}
\newcommand\bpm{\begin{figure*}[!ht]}
\newcommand\epm{\end{figure*}}

\begin{document}

\title{The \emph{Fermi} Gamma-Ray Haze from \\
       Dark Matter Annihilations and Anisotropic Diffusion}

\author{Gregory Dobler\altaffilmark{1,5},
  Ilias Cholis\altaffilmark{2,3,6}, 
  \& Neal Weiner\altaffilmark{3,4,7}}

\altaffiltext{1}{
  Kavli Institute for Theoretical Physics,
  University of California, Santa Barbara
  Kohn Hall, Santa Barbara, CA 93106 USA
}
\altaffiltext{2}{
  Astrophysics Sector,
  La Scuola Internazionale Superiore di Studi Avanzati and 
  Istituto Nazionale di Fisica Nucleare, Sezione di Trieste,
  via Bonomea 265, 34136 Trieste, Italy }
\altaffiltext{3}{
  Center for Cosmology and Particle Physics,
  Department of Physics,
  New York University, New York, NY 10003 USA }
\altaffiltext{4}{School of Natural Sciences, Institute for
Advanced Study, Princeton, NJ 08540}

\altaffiltext{5}{dobler@kitp.ucsb.edu}
\altaffiltext{6}{ilias.cholis@sissa.it}
\altaffiltext{7}{neal.weiner@nyu.edu}

%%%%%%%%%%%%%%%%
%%  Abstract  %%
%%%%%%%%%%%%%%%%

\begin{abstract}
  Recent full-sky maps of the Galaxy from the \emph{Fermi Gamma-Ray
    Space Telescope} have revealed a diffuse component of emission
  towards the Galactic center and extending up to roughly $\pm$50
  degrees in latitude.  This \emph{Fermi} ``haze'' is the inverse
  Compton emission generated by the same electrons which generate the
  microwave synchrotron haze at WMAP wavelengths.  The gamma-ray haze
  has two distinct characteristics: the spectrum is significantly
  harder than emission elsewhere in the Galaxy and the morphology is
  elongated in latitude with respect to longitude with an axis ratio
  $\approx$2.  If these electrons are generated through annihilations
  of dark matter particles in the Galactic halo, this morphology is
  difficult to realize with a standard spherical halo and isotropic
  cosmic-ray diffusion.  However, we show that anisotropic diffusion
  along ordered magnetic field lines towards the center of the Galaxy
  coupled with a prolate dark matter halo can easily yield the
  required morphology without making unrealistic assumptions about
  diffusion parameters.  Furthermore, a Sommerfeld enhancement to the
  self annihilation cross-section of $\sim$30 yields a good fit to the
  morphology, amplitude, and spectrum of both the gamma-ray and
  microwave haze.  The model is also consistent with local cosmic-ray
  measurements as well as CMB constraints.
\end{abstract}

%%%%%%%%%%%%%%%%%%%%
%%  Introduction  %%
%%%%%%%%%%%%%%%%%%%%

\section{Introduction}
\label{sec:introduction}

With the first year data release, the \emph{Fermi Gamma-Ray Space
  Telescope} provided a wealth of new insights and detail of the
gamma-ray sky.  The energy range and angular resolution of the Large
Area Telescope (LAT) on board \emph{Fermi} has significantly advanced
the understanding of many areas of gamma-ray astronomy, from point
source studies like pulsars \citep{2009Sci...325..848A,
  2010ApJS..187..460A, 2010arXiv1006.2134S} and blazars
\citep{2009ApJ...700..597A, 2010ApJ...715..429A, 2010ApJ...716...30A,
  2010ApJ...722..520A}, to diffuse emissions from the extragalactic
gamma-ray background \citep{Abdo:2010nz, 2010ApJ...717L..71A,
  Abdo:2010dk} and the interstellar medium (ISM) \citep{Abdo:2009ka,
  Porter:2009sg, 2010ApJ...722L..58S}.

Recently, \cite{dobler:2010fh} assembled full-sky maps of the Galaxy
using the published raw photon data from \emph{Fermi} from several
hundred MeV up to several hundred GeV.  These maps of gamma-ray
emission from the diffuse ISM are produced primarily through three
processes: cosmic-ray (CR) protons collide with the ISM producing
$\pi^0$ particles that decay to gammas, bremsstrahlung from CR
electrons (and positrons) colliding with ions, and inverse Compton
(IC) scattering of starlight, infrared, and CMB photons by CR
electrons.  Because bremsstrahlung and $\pi^0$ emission are due to
collisions of CRs with the ISM, these emissions are highly spatially
correlated with other maps of the interstellar medium like the dust
column density map of \cite{1998ApJ...500..525S}.  Since the IC
emission is generated by interactions of CR electrons with the
interstellar radiation field (ISRF) there is not a good morphological
tracer of this emission at other energies.  However, CR electrons are
primarily accelerated in supernova (SN) remnants and so their
injection morphology should be very disk-like.  Although diffusion
effects are important, for isotropic diffusion through the Galaxy the
resultant IC emission should also be very disk-like.

Using template fitting techniques to morphologically regress out the
emission from $\pi^0$'s, bremsstrahlung, and IC from disk electrons
from the \emph{Fermi} maps, \cite{dobler:2010fh} found an excess
``haze'' of IC emission towards the Galactic center (GC) extending
$\pm$50 degrees in latitude and with an axis ratio of roughly 2.0.
This \emph{Fermi} haze is the gamma-ray counterpart to the microwave
haze observed by the \emph{Wilkinson Microwave Anisotropy Probe}
(WMAP) as described in \cite{Finkbeiner:2003im} and
\cite{Dobler:2007wv}.  At WMAP wavelengths, the same electrons which
generate the \emph{Fermi} IC haze interact with the Galactic magnetic
field to produce synchrotron microwaves.  Recently \cite{Su:2010qj}
reconsidered the morphology, arguing for a ``bubble''-like
structure. Nonetheless, for reasons outlined in
\refsec{haze_morphology} we use the ``haze'' moniker throughout this
paper, although we are considering effectively the same gamma-ray
signal.

In both the gamma-ray and synchrotron cases, the haze emission is
significantly harder than elsewhere in the Galaxy, implying that the
electrons which produce the haze have a harder spectrum than the
electrons accelerated and diffused through the Galactic disk.  In
fact, the required electron spectrum (number density per unit energy)
is roughly $dN/dE \propto E^{-1}$ at high energies which is
significantly harder than electrons generated by SN shock acceleration
after taking into account diffusion effects.  In that case, the steady
state spectrum is closer to $dN/dE \propto E^{-3}$.

The identification of the haze in both the WMAP and \emph{Fermi} data
imply that the haze is both real and that the underlying electron
spectrum is very hard.  It is this hard spectrum and the diffuse
elongated morphology that are the defining characteristics of the
emission, and any proposed origin for the electrons must match both of
these features.  For example, several authors have studied the
connection between the haze electrons and young and middle aged
pulsars \citep{Zhang:2008tb, FaucherGiguere:2009df, McQuinn:2010ju}.
The morphology however of the diffused electrons accelerated in pulsar
winds would also be very disk-like and would not match the
morphology\footnote{Millisecond pulsars in the galactic halo may
  contribute to the haze signal at some level
  \citep[see][]{2010ApJ...722.1939M}, but their morphology would also
  likely be spherical instead of significantly elongated in
  latitude.}.  Others have tried to reproduce the haze emission with a
combination of increased SN rate and modified diffusion parameters
\citep{McQuinn:2010ju, Gebauer:2009hk}, but this also cannot produce
the observed morphology or the observed spectrum, even including
possible reacceleration effects.  Lastly, there has been speculation
that both the gamma-ray haze \citep{Linden:2010ea} and the microwave
haze \citep{Mertsch:2010ga} are due to imperfect template subtraction,
however neither of these criticisms has been able to produce the
morphology or the spectrum (amplitude and shape) of the observations
using simulations.  Furthermore, the gamma-ray haze is visible in the
\emph{Fermi} sky maps \emph{without performing any template fitting}
demonstrating that it is clearly a real structure.

This work builds upon previous studies of the haze which explore the
possibility that the haze electrons are generated through dark matter
(DM) annihilations in the Galactic halo.  \cite{finkbeiner:2004us}
originally showed that the microwave haze morphology and spectrum in
the WMAP 1-year data was reasonably well matched by a DM model with a
particle mass of $M_{\chi} \sim 100$ GeV and with a self annihilation
cross-section $\sigmav \sim 3 \times 10^{-26}$ cm$^3/$s which is
roughly that required to yield the observed relic density of DM
$\Omega_{DM} \approx 0.23$ if the DM particle is a thermal relic of
the Big Bang.

However, initial data from \emph{Fermi} of the inner Galaxy suggested
that the IC emission from the haze electrons extended up to at least
$\sim200$ GeV implying a DM particle mass of closer to $\sim1$ TeV.
Since the annihilation rate is proportional to the number density
squared, this requires a $\sigmav$ roughly 100 times the thermal relic
value in order to match the data. With light force carriers, a ``boost
factor'' of 100 in the Galactic halo is easily obtainable
\citep{ArkaniHamed:2008qn,Pospelov:2008jd} via the Sommerfeld
mechanism \citep{sommerfeld, Hisano:2004ds,
  Hisano:2003ec,Cirelli:2008id,Lattanzi:2008qa}, in which $\sigmav$
increases with decreasing relative velocity up to some saturation
value, while still producing the correct relic density
\citep{Feng:2010zp, Finkbeiner:2010sm}.  Such a particle model is also
consistent with local electron and positron CR anomalies observed by
the \emph{Payload for Antimatter Exploration and Light-nuclei
  Astrophysics} \citep[PAMELA;][]{Picozza:2006nm, Adriani:2008zr,
  Adriani:2010ib} satellite and \emph{Fermi} \citep{Abdo:2009zk,
  Ackermann:2010ij} as shown by \cite{Cholis:2008wq} and
\cite{Cholis:2009va}.  A model independent fit to all of the data
(gammas, microwaves, and CRs) by \cite{2010PhRvD..82b3518L} confirms
that the injection spectrum must be $E^2dN/dE \propto E^2$ which is
broadly consistent with the spectrum of a Sommerfeld enhanced DM
annihilation scenario in which the main products are leptons.

These works have shown that the amplitude and spectrum of the haze are
easily reproduced with a DM particle annihilation model; but here we
are concerned primarily with the morphology.  The morphology of the
gamma-ray haze is the most difficult aspect to model since the haze is
significantly elongated in latitude with respect to longitude.  In
fact, the geometry is impossible to realize with disk-like (or, as we
show in \refsec{results}, spherical) injection, ruling out SNe or
pulsars as a possible source.

Such a geometry is also inconsistent with a spherical DM halo and
isotropic diffusion.  However, it is very likely that neither of these
assumptions is accurate.  Generically, DM N-body simulations of Milky
Way sized halos imply prolate halos with an axis ratio of roughly 2
\citep{2008Natur.454..735D, 2008JPhCS.125a2008K, 2008MNRAS.391.1685S}
and observations of the spatial distribution of Milky Way satellites
imply a prolate halo oriented perpendicular to the Galactic disk
\citep[e.g.,][]{zentner:2005ad}.  In addition, the presence of any
ordered magnetic field lines towards the GC implies that the electrons
will not diffuse isotropically as they follow the fields.  In
\refsec{haze_morphology} we discuss the morphology of the haze in more
detail, and in \refsec{diffmodel} we outline our anisotropic diffusion
model which produces a DM IC halo that closely resembles the observed
morphology.  In \refsec{results}, we compare our model to the data
(both the morphology, amplitude, and spectrum of the haze emission)
and in \refsec{conclusions} we summarize our conclusions.

%%%%%%%%%%%%%%%%%%%%%%%
%%  Haze morphology  %%
%%%%%%%%%%%%%%%%%%%%%%%

\section{Haze morphology}
\label{sec:haze_morphology}

Prior to the release of the gamma-ray data, the microwave haze was
described by \cite{Finkbeiner:2003im} and \cite{Dobler:2007wv} as
being centered on the GC, roughly spherical, and decreasing in
amplitude approximately as $1/r$ where $r$ is the angular distance to
the GC.  However, such a microwave signal is limited by the extent of
the B-field off the disk.  The \emph{Fermi} data on the other hand
clearly show that the haze is in fact elongated in latitude $b$ and
extends to $|b|\sim50$ degrees.  Despite the lower angular resolution
and signal-to-noise, the gamma-ray data give a more complete picture
of the location of the haze electrons.  The reason for the different
morphologies is that the synchrotron amplitude is proportional to the
magnetic field strength while the IC is proportional to the ISRF.
Since the magnetic field falls off quickly with distance above the
Galactic disk while the CMB amplitude is latitude independent, the
microwave haze is confined to lower latitudes compared to the
gamma-ray haze.

\bpm
  \centerline{
    \includegraphics[width=0.49\textwidth]{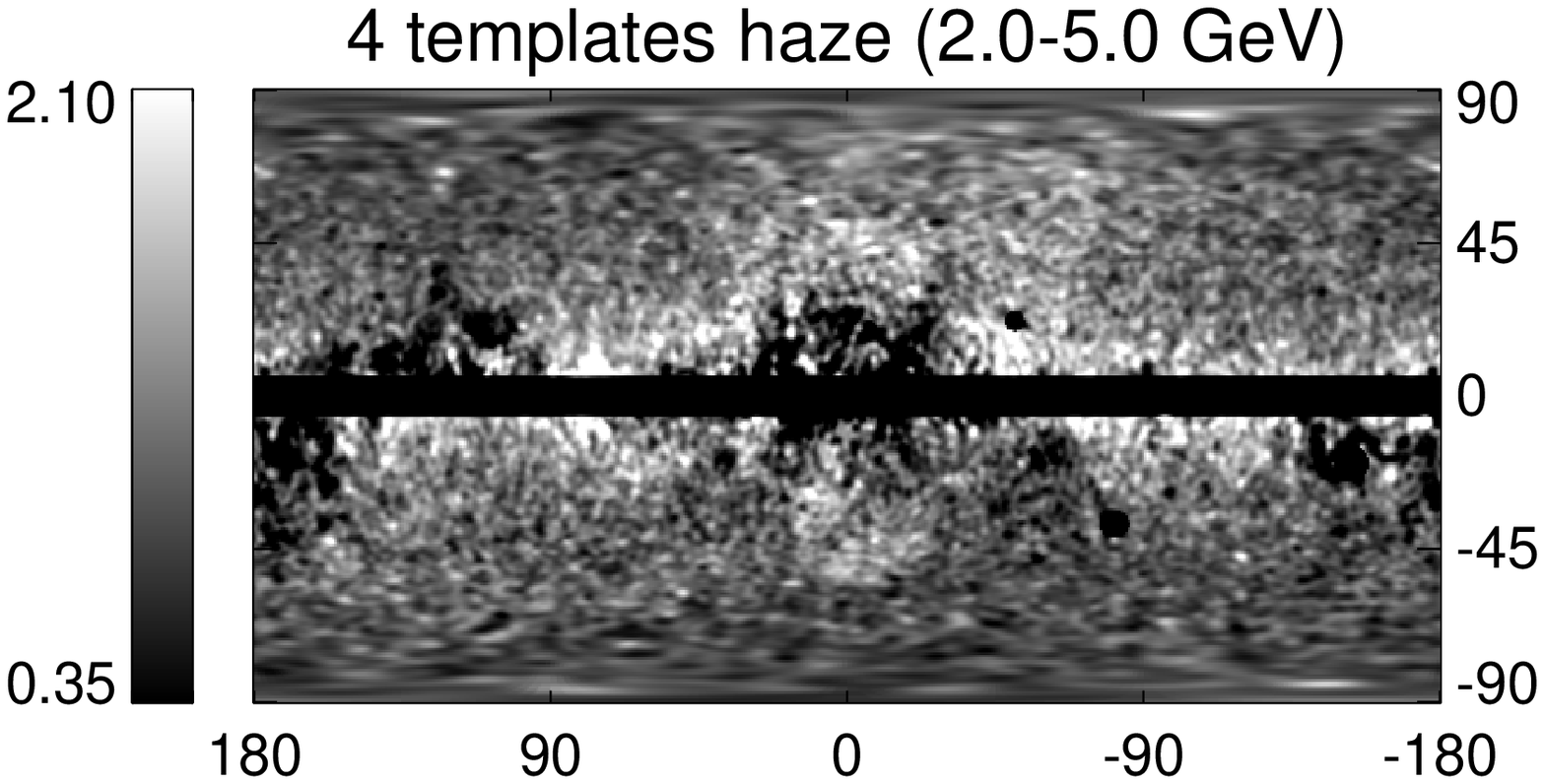}
    \includegraphics[width=0.49\textwidth]{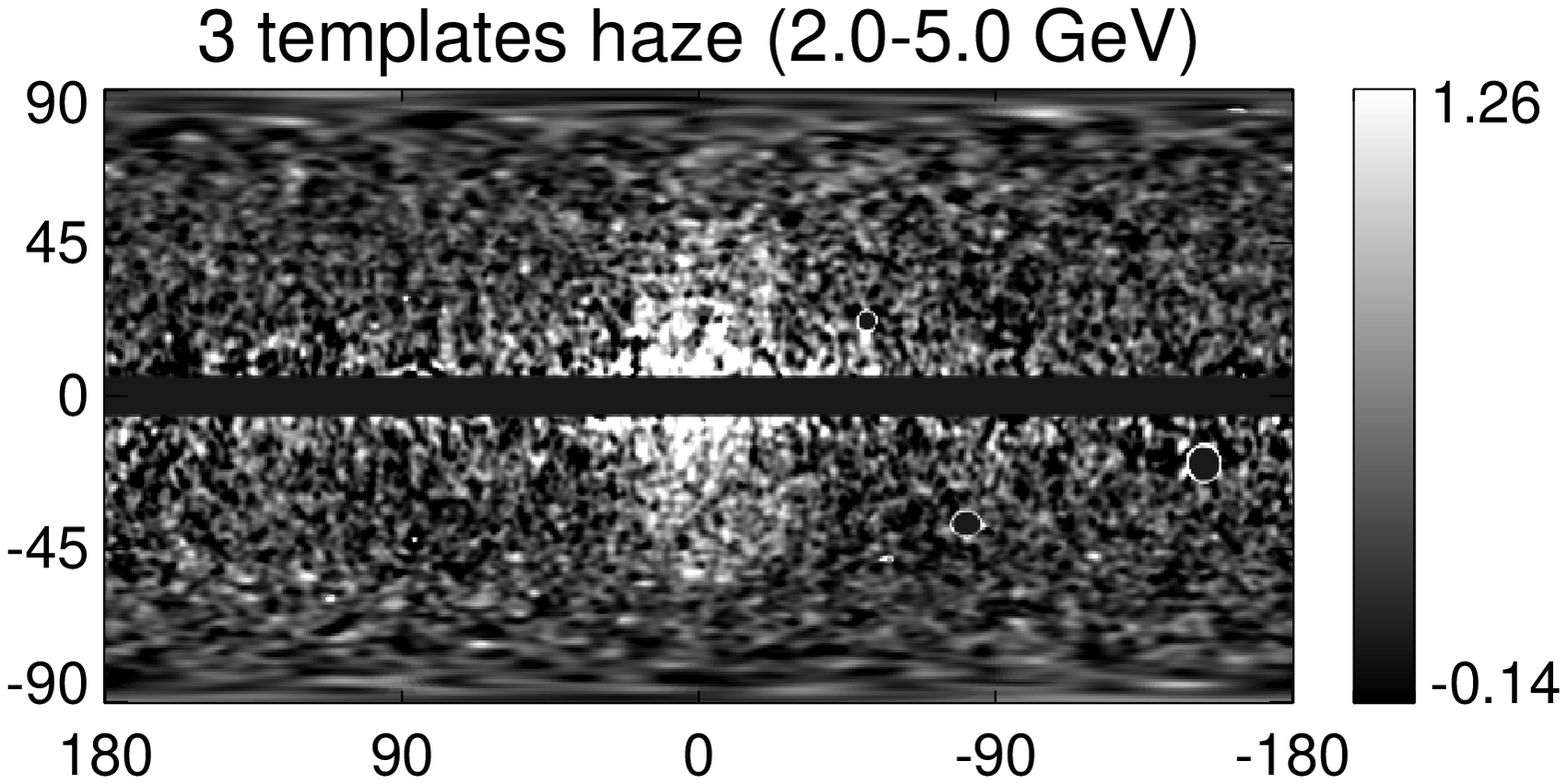}
  }
  \centerline{
    \includegraphics[width=0.49\textwidth]{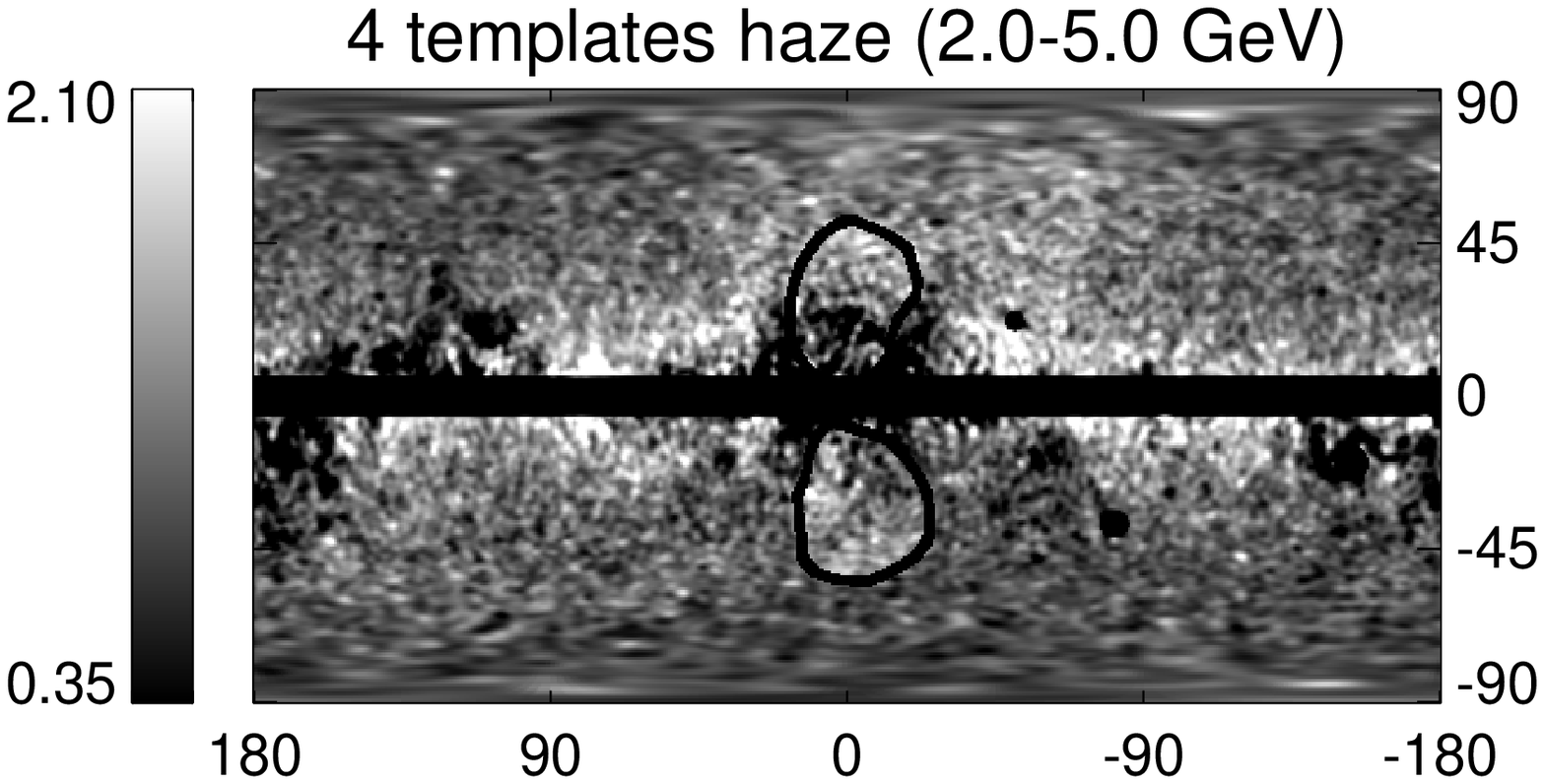}
    \includegraphics[width=0.49\textwidth]{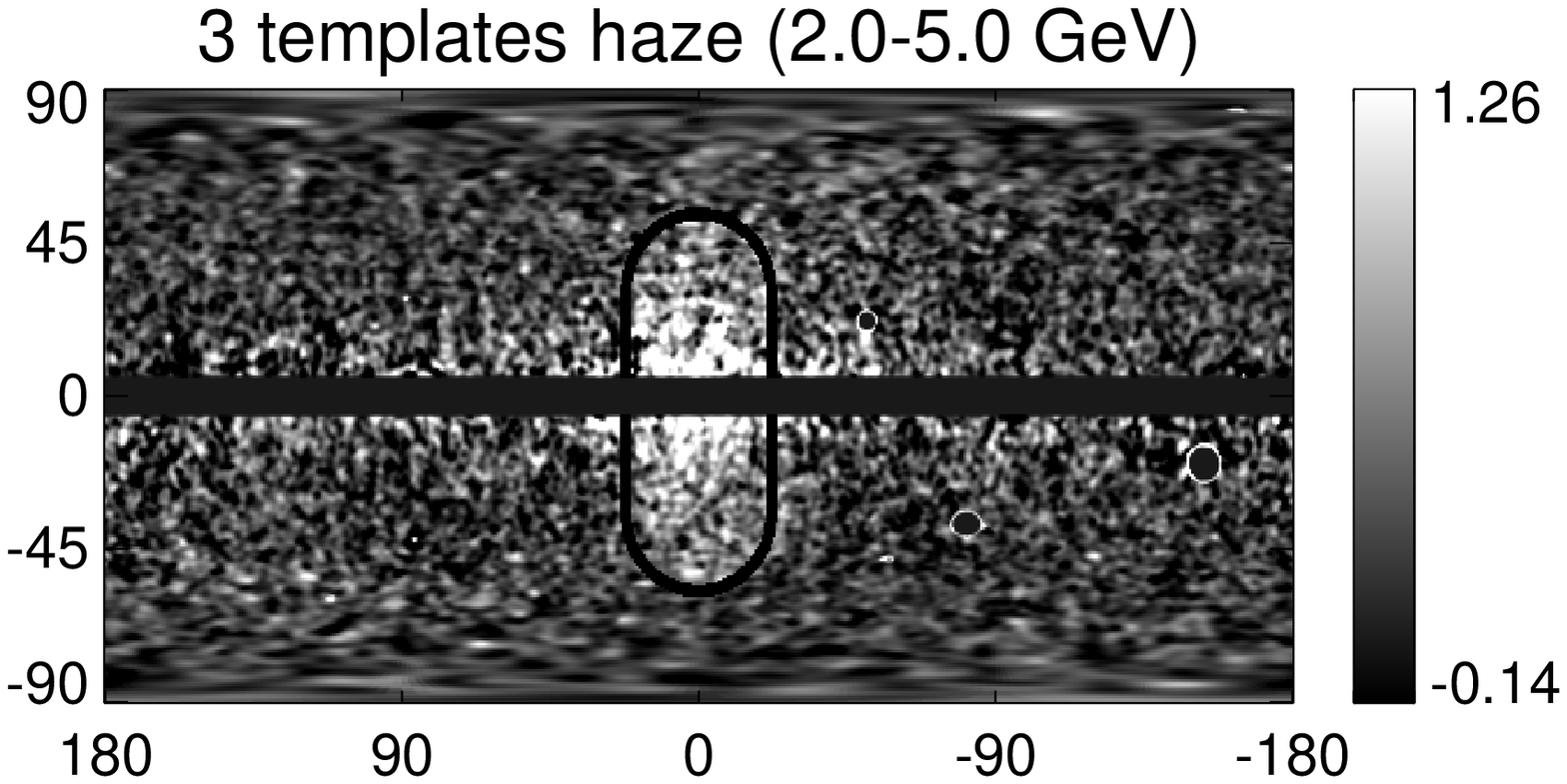}
  }
\caption{
  \emph{Upper left:} The haze residual using the 4 templates fit
  (disk, SFD, uniform, bubble) defined in \cite{Su:2010qj}.  The haze
  residual in this case is very pinched in the center and resembles
  two ``bubbles''; however, note the significant regions of
  over-subtraction near the disk which pinch the haze towards the
  center.  \emph{Upper right:} The haze residual using the 3 templates
  fit ($E_{0.5}^{1.0}$, uniform, GALPROP).  Although there is more
  noise, there is now very little disk over-subtraction and the haze
  looks much more like an ``oval''.  \emph{Bottom row:} the same
  residuals with hand drawn contours over-plotted to highlight the
  morphological differences.
}
\label{fig:compmorph}
\epm

The detailed morphology of the gamma-ray haze close to the Galactic
plane is difficult to determine.  In \cite{dobler:2010fh}, three
methods of template fitting were used: 1) the actual \emph{Fermi} data
from 1.0-2.0 GeV was used as a full-sky template, 2) the
\cite{1998ApJ...500..525S} (SFD) dust map was used alone, and 3) the
SFD dust map, the Haslam 408 MHz map \citep{1982A&AS...47....1H}, and
a bivariate Gaussian haze template were used.  Method 2) was not
particularly successful at fitting the full sky data and left
significant disk-like residuals as well as the \emph{Fermi} haze.
Method 1) and 3) were much more successful but gave very different
haze morphologies at low latitudes ($|b|<30$ deg).  In particular,
using method 1) gives a haze which is more oval shaped while method 3)
gives a haze which is more hourglass or ``bubble'' shaped (see
\reffig{compmorph}).  Recently, \cite{Su:2010qj} explored the bubble
morphology of method 3) in detail and argued that this morphology may
be indicative of a significant event towards the GC (e.g., accretion
onto the central black hole) in the past.  However, before ascribing a
physical mechanism to the generation of the haze electrons which is
dependent upon the haze morphology, it is important to determine what
that morphology \emph{is} and why the two methods differ.

Both methods 1) and 3) have associated problems.  Since method 1)
takes differences of \emph{Fermi} data at different energies, any haze
that is present in the lower energy data is subtracted off of the
higher energy data so that the specific spectrum of the \emph{Fermi}
haze cannot be uniquely determined.  In addition, since the
\emph{Fermi} maps have somewhat low signal to noise, subtracting one
map from another (which adds the noise in weighted quadrature while
removing the signal) yields difference maps that can be quite noisy.

On the other hand, method 1) has the advantage that it does not rely
on external templates (like the SFD dust map for example) and so
automatically takes into account systematics like line of density
effects in the ISM.  In other words, the lower energy \emph{Fermi}
maps are a better morphological tracer of the higher energy
\emph{Fermi} maps than external templates.  The fact that the haze
residual remains in the difference is a statement that this emission
has a significantly harder spectrum than the emission elsewhere in the
Galaxy.

\bpm
  \centerline{
    \includegraphics[width=0.49\textwidth]{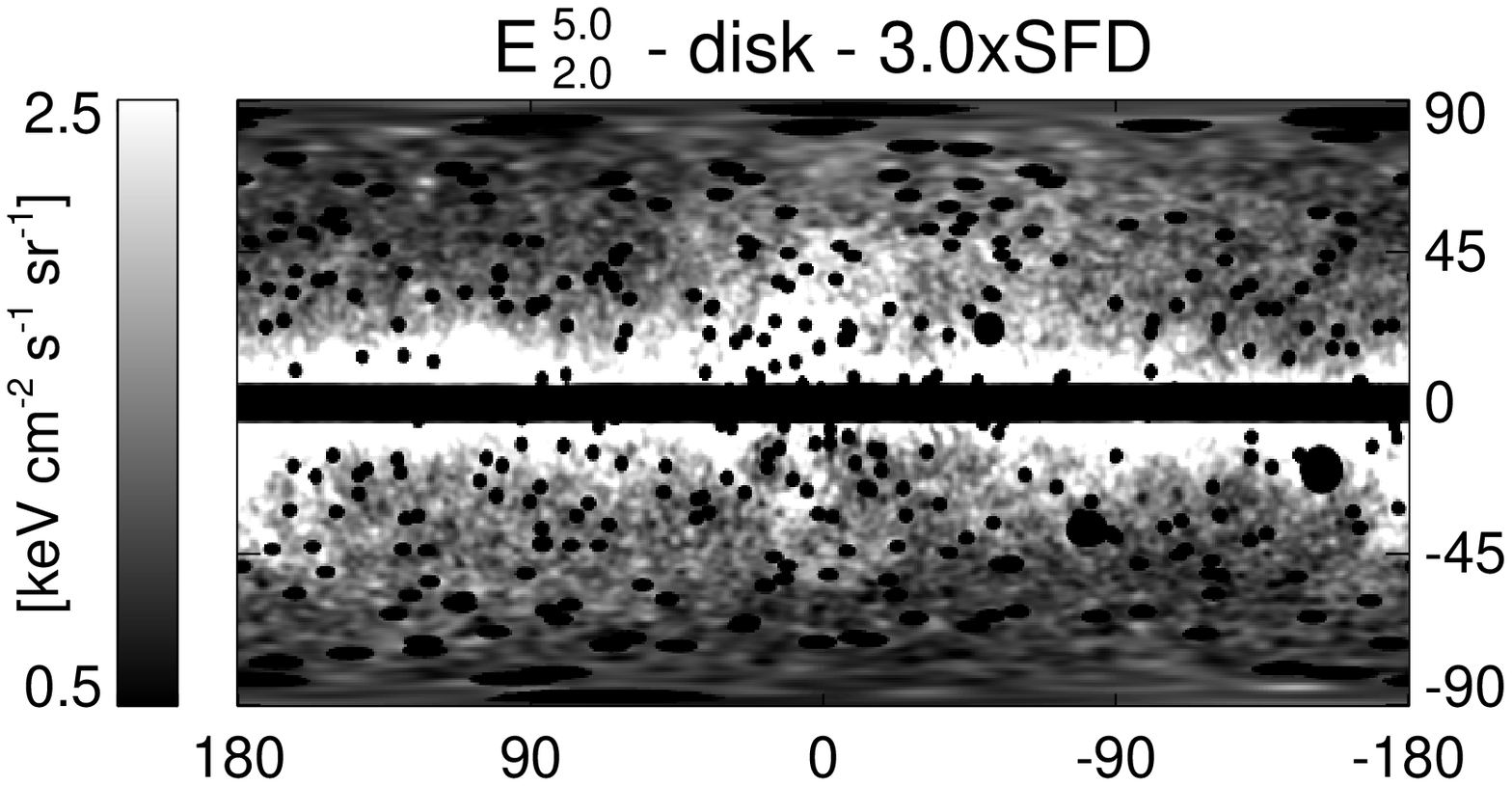}
    \includegraphics[width=0.49\textwidth]{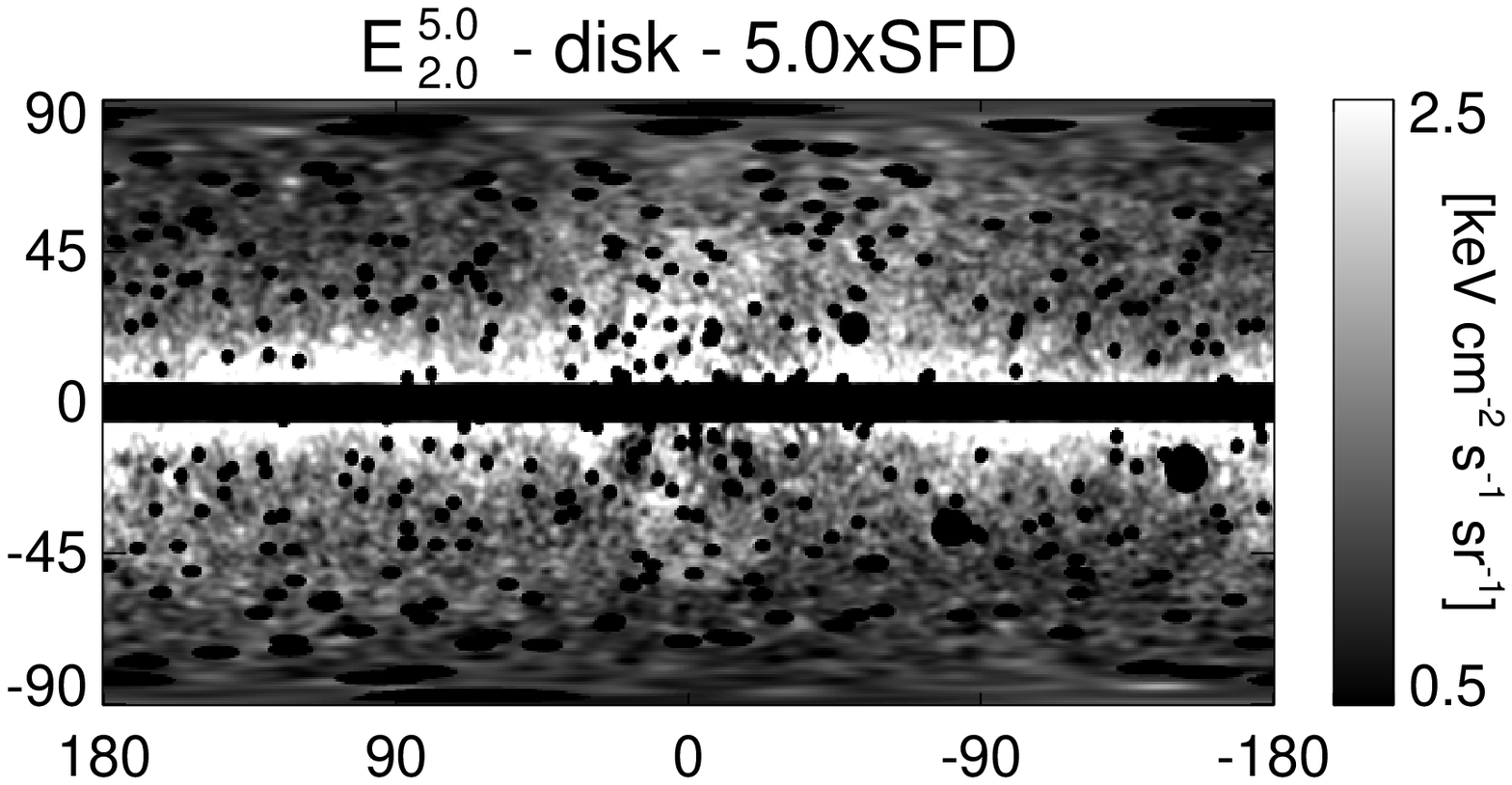}
  }
  \centerline{
    \includegraphics[width=0.49\textwidth]{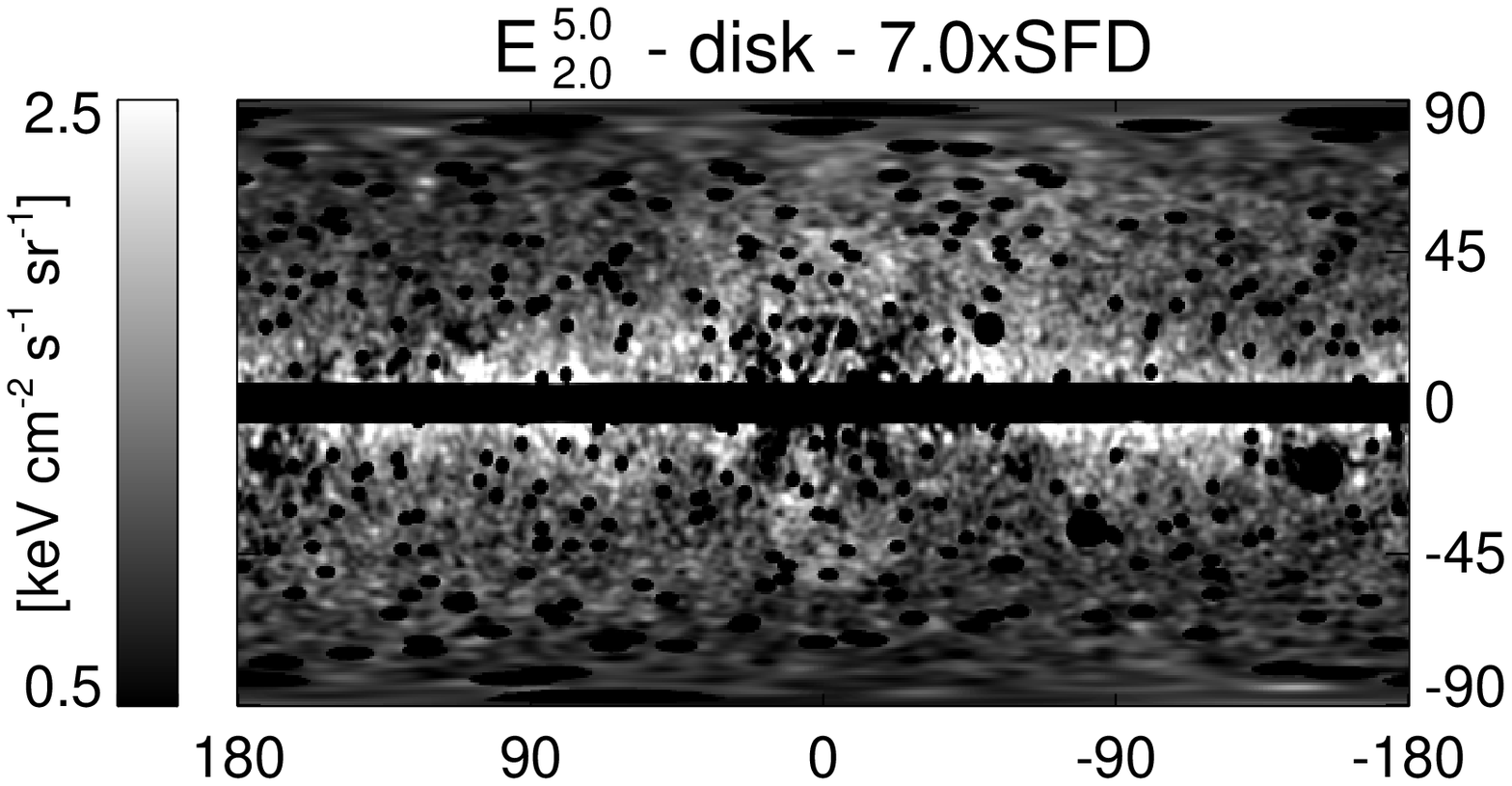}
    \includegraphics[width=0.49\textwidth]{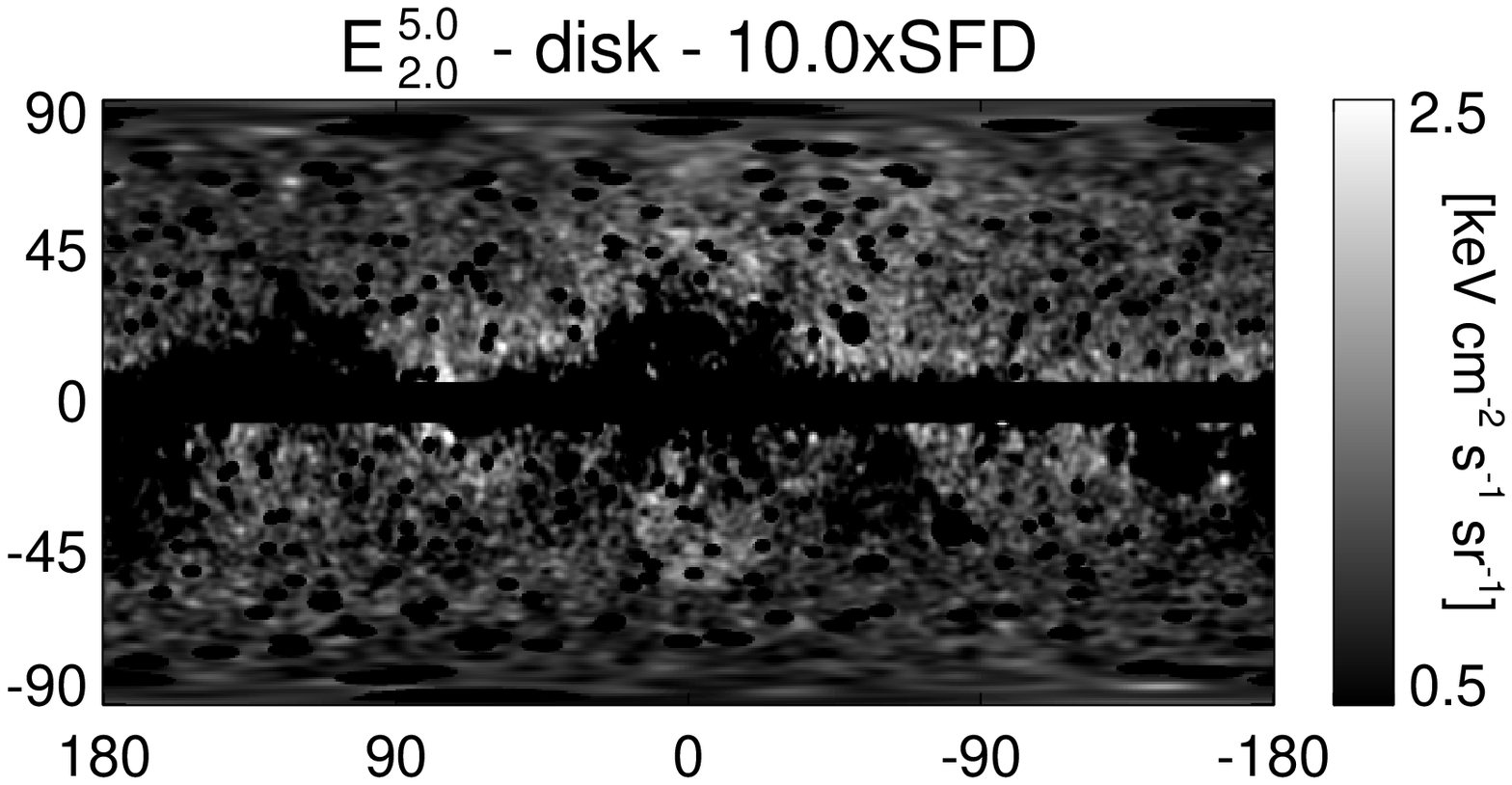}
  }
\caption{
  The \emph{Fermi} data at 2.0-5.0 GeV minus the disk IC model of
  \cite{Su:2010qj} plus varying amplitudes times the SFD dust map as a
  tracer of $\pi^0$ emission.  As the SFD amplitude is increased a
  clear ``X'' shape or over-subtraction emerges towards the Galactic
  center.  This is due to lower $\pi^0$ to dust column ratio in that
  shape towards the bulge likely caused by line of sight density
  variations of the ISM and cosmic ray protons.  This X-shaped
  over-subtraction can make the oval shaped haze (right hand panel of
  \reffig{compmorph}) appear more ``bubble''-shaped (left hand panel
  of \reffig{compmorph}).
}
\label{fig:Xbulge}
\epm

The advantage of method 3) is that the absolute spectrum of the haze
can be well determined since the haze structure is not in the external
templates.  However, because of line of sight variations in the ISM
and cosmic-ray proton density, there will be, for example, variations
in the ratio of $\pi^0$ gamma-ray emissivity to total dust column
density.  Thus the dust column map will not be a perfect tracer of the
gamma-ray map.  This is especially true in the inner Galaxy (within
about 30 deg of the GC) and has the potential to significantly effect
the perceived haze morphology.  To illustrate this point,
\reffig{Xbulge} shows the \emph{Fermi} data from 2.0-5.0 GeV with the
\cite{Su:2010qj} model for IC emission and varying amounts of the SFD
map subtracted.  When the SFD coefficient is small, the $\pi^0$ gammas
are clearly under-subtracted.  However, as the coefficient is
increased, a clear ``X'' shaped over-subtraction becomes visible.
This structure defines the ``bubble'' shape of the haze in method 3),
and may be the root of the discrepancy between the two morphologies.
That is, if the haze were actually oval shaped, it may appear more
hourglass shaped after over-subtracting this ``X''.

It is important to note that this ``X'' is not a feature \emph{in} the
SFD map (with the exception of the upper-right and possibly
lower-right edges) but rather is being over-subtracted because the
projected $\pi^0$ to dust column ratio is lower in that shape.

Furthermore, it is quite possible that the environmental conditions
towards the GC which give rise to this ``X'' in gammas, produce
similar features in X-rays and microwaves.  For example, a heating
source towards the center could heat the gas leading to enhanced,
harder x-ray emission and such a variation in the environment would
affect the estimate of column density to spinning dust emissivity used
by \cite{Dobler:2007wv,dobler:2008sd} and \cite{Dobler:2008av} to
remove the spinning dust component at microwaves.  This would have the
affect of making both the gamma-ray and microwave haze more hourglass
shaped due to the same ISM physics which generates an edge in x-rays.
Without speculating further what this ``X'' structure is, we note that
there is significant evidence for X-shaped bulges in other galaxies,
and recent evidence from the 2MASS survey that there exist red clump
populations in the Milky Way that follow this feature
\citep{mcwilliam:2010rc}.

In the context of comparing the gamma-ray haze spectrum and morphology
to a signal generated by injecting electrons via dark matter
annihilations, the ``bubble'' morphology seems difficult to obtain (or
at the very least, seems more indicative of a transient event in the
GC).  However, we show below that an oval shaped haze (and even an
hourglass shaped haze) is possible with DM annihilation when
considering anisotropic diffusion effects.  Regardless, the underlying
morphology of the gamma-ray haze at low latitudes is an unsettled
issue.  We choose to compare our results to the oval-shaped morphology
and show that method 1) plus a dark matter contribution to the IC
emission with anisotropic diffusion effects is consistent with the
data.

%%%%%%%%%%%%%%%%%%%%%%%
%%  Diffusion model  %%
%%%%%%%%%%%%%%%%%%%%%%%

\section{Diffusion model}
\label{sec:diffmodel}

Since the basis for any anisotropic diffusion scenario is that
electrons travel along ordered field lines, our diffusion model must
first assume a geometry for the ordered component of the Galactic
magnetic field.  From there, this magnetic field can be related to
specific diffusion parameters which appear in the diffusion equation.
All of our calculations are done by modifying the CR propagation code
GALPROP \citep{1998ApJ...509..212S, 2001AdSpR..27..717S,
  2003ICRC....4.1917M, 2006ApJ...642..902P, Strong:2007nh} to include
anisotropic effects.

%%%%%%%%%%%%%%%%%%%%%%%%%%%%%%%%%%%%%
%%  Galactic Magnetic Field Model  %%
%%%%%%%%%%%%%%%%%%%%%%%%%%%%%%%%%%%%%

\subsection{Galactic magnetic field model}

Our magnetic field model consists of two components: an irregular
magnetic field $B_{\rm irr}$ and an ordered magnetic field $B_{\rm
  ord}$.  The former is parameterized as an exponential disk,
\beq
  B_{\rm irr} = B_0 e^{(R_{\odot}-r)/r_1-|z|/z_1},
  \label{eq:bfieldirr}
\eeq
where $r$ and $z$ are the radial and vertical distances from the GC
respectively, and $B_0$ is the local value of the irregular component
(i.e., at $r = R_{\odot} \approx 8.5$ kpc, the GC-sun distance).  The
ordered field is assumed to have the form,
\beq
  B_{\rm ord} = B_1 e^{-r/r_2 - |z|/z_2} \times \left(1 + K
  e^{-r/r_{3} - |z|/z_3} \right),
  \label{eq:bfieldord}
\eeq
where $B_1 (1+K)$ is the amplitude of the ordered field at the GC,
which is based on the 3D field model of \cite{2010HEAD...11.3206O}.

\bp
  \centerline{
    \includegraphics[width=0.49\textwidth]{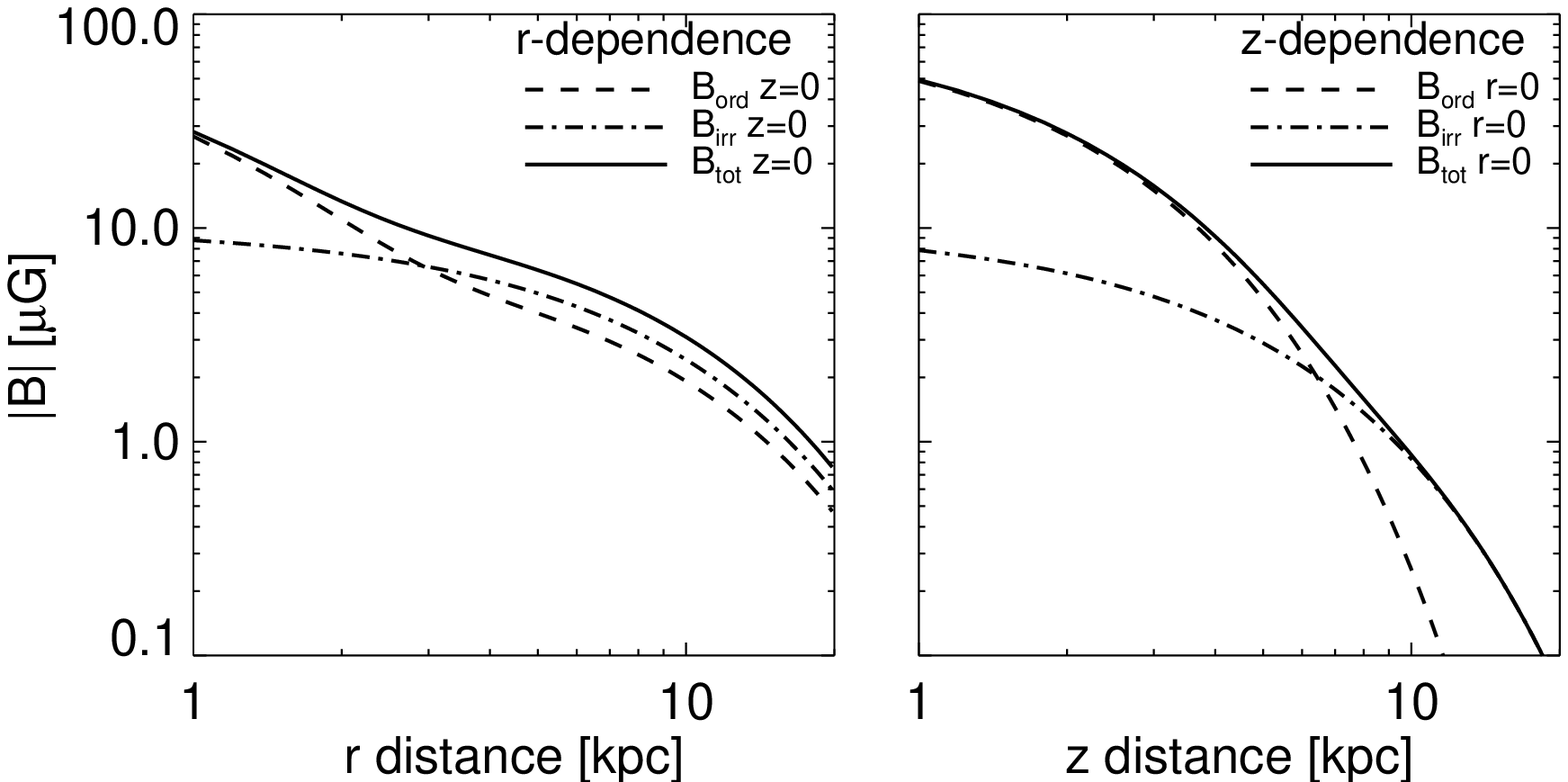}
  }
\caption{
  B-field amplitude profiles versus radial distance $r$ (\emph{left})
  and versus distance above the galactic disk $z$ (\emph{right}).
}
\label{fig:BfieldProfiles}
\ep

\begin{deluxetable*}{|c|c|c|c|c|c|c|c|c|c|c|}
  \tablehead{
    Model & $B_{\rm ord}$ Formula & $B_0$ & $r_1$ & $z_1$ & $B_1$ &
    $K$ & $r_2$ & $z_2$ & $r_3$ & $z_3$ \\
    & & ($\mu$G) & (kpc) & (kpc) & ($\mu$G) & & (kpc) & (kpc) & (kpc)
    & (kpc)
  }
  \startdata
    1 & $B_1 e^{-r/r_2-|z|/z_2} \times
    \left(1+Ke^{-r/r_3-|z|/z_3}\right)$ & 3 & 7 & 4 & 8 & 10 & 7 & 2 &
    0.8 & 10 \\
    2 & $B_1 e^{-r/r_2-|z|/z_2} \times \left(1+Ke^{-(r/r_3)^2}
    \sqrt{\cos(|z|/z_3\times\pi/2)}\right)$ & 3 & 5 & 4 & 10 & 11 & 5
    & 4 & 1 & 40 \\
    3 & $B_1 e^{-r/r_2-|z|/z_2} \times
    \left(1+Ke^{-(r/r_3)^{1.5}-|z|/z_3}\right)$ & 3 & 10 & 2 & 10 & 6
    & 10 & 3 & 1.2 & 20 \\
    4 & $B_1 e^{-r/r_2-|z|/z_2} \times
    \left(1+Ke^{-(r/r_3)^{1.5}-(|z|/z_3)^{1.5}}\right)$ & 3.7 & 5 & 2
    & 12.5 & 8 & 7 & 5 & 2.5 & 20 \\
    5 & $B_1 e^{-r/r_2-|z|/z_2} \times
    \left(1+Ke^{-r/r_3-|z|/z_3}\right)$ & 3.7 & 5 & 2 & 3.7 & 12 & 5 &
    2 & 2 & 6
  \enddata
  \tablecomments{
    Magnetic field morphologies and parameters for the IC signals
    plotted in \reffig{varyBfieldmorph}.  Our fiducial model is Model
    1 which generates an IC signal that roughly matches the
    \emph{Fermi} haze morphology (see \reffig{iccomponents}).
}\label{tbl:Bfieldmods}
\end{deluxetable*}

The parameters $B_0$, $B_1$, $K$, $r_{1,2,3}$, and $z_{1,2,3}$ that we
use are set by hand to reproduce the appropriate IC geometry and agree
with measured values of the Galactic magnetic field at distances
greater than $\sim1$ kpc from the GC \citep[][see
  \reffig{BfieldProfiles}]{Jansson:2009ip, 1983ApJ...265..722S,
  1994A&A...288..759H, Beck:2000dc, Han:2001vf, Tinyakov:2001ir,
  Sun:2007mx, Brown:2007qv, beck:2009mm, 2010MNRAS.401.1013J,
  2010ApJ...722L..23N}.  The parameters of our fiducial model (Model
1) are shown in Table \ref{tbl:Bfieldmods}.  These give a local value
for the total magnetic field of 5.4 $\mu$G and an ordered-to-total
amplitude ratio of $\approx$0.62 which agrees well with measured
values \citep[see][and references
  therein]{beck:2009mm}.\footnote{These parameters do give a somewhat
  high value of 89 $\mu$G for the total field at the very center,
  $r=z=0$ kpc.  However, we note that not only is this in agreement
  with the estimates of \cite{2010Natur.463...65C} who place a
  \emph{lower} limit of 50 $\mu$G in the inner 400pc from necessary
  synchrotron cooling to avoid violating existing diffuse $\gamma$-ray
  bounds, but also the very center is well outside our region of
  interest.  Our mask of the Galactic plane extends up to $|b|=5$ deg
  or $|z| \approx 0.75$ kpc.  Inside this region, our choice of
  B-field has little impact on our results and our value at the center
  is only due to our specific parameterization of the field which
  likely does not extend in to arbitrarily small distances.}

%%%%%%%%%%%%%%%%%%%%%%%%%%%%%
%%  Anisotropic diffusion  %%
%%%%%%%%%%%%%%%%%%%%%%%%%%%%%

\subsection{Anisotropic diffusion}
\label{sec:anisodiff}

The propagation of CRs through the ISM is governed by the diffusion
equation,
\beq
  \frac{\partial \psi}{\partial t} = \frac{\partial (b\psi)}{\partial
    E} + \overrightarrow{\nabla}(D\overrightarrow{\nabla} \psi) + Q,
  \label{eq:Propagation_equation}
\eeq
where $\psi$ is the number density per unit particle momentum of CRs
at time $t$ and position $\vec{x}$, $b$ is an energy loss coefficient
(dominated by synchrotron and IC in the case of electron CRs), $Q$ is
a source term due to the injection of electrons by DM annihilations,
and $D$ is the diffusion constant.  It is this last parameter which
must be modified for the case of anisotropic diffusion, and so we are
concerned with the $\overrightarrow{\nabla}(D\overrightarrow{\nabla}
\psi)$ term above.

We solve Equation \ref{eq:Propagation_equation} using GALPROP on a
cylindrical grid so that,
\beq
    \overrightarrow{\nabla}(D\overrightarrow{\nabla} \psi) =
    \frac{1}{r}\frac{\partial}{\partial r}(r D
    \frac{\partial\psi}{\partial r}) + \frac{\partial}{\partial z}(D
    \frac{\partial\psi}{\partial z}).
\label{eq:Diffusion_term}
\eeq

Typically, isotropic diffusion is assumed so that $D$ is not a
function of $\vec{x} = (r,z)$.  However in our case
Eq.\ \ref{eq:Diffusion_term} generalizes to:
\begin{eqnarray}
  \overrightarrow{\nabla}(D\overrightarrow{\nabla} \psi) &=&
  \frac{1}{r}\frac{\partial}{\partial r}(r D_{rr}
  \frac{\partial\psi}{\partial r} + r
  D_{rz}\frac{\partial\psi}{\partial z}) \nonumber \\
  &+& \frac{\partial}{\partial z}(D_{zz} \frac{\partial\psi}{\partial
    z} + D_{zr}\frac{\partial\psi}{\partial r}),
\label{eq:Diffusion_termGeneral}
\end{eqnarray}
where $D_{rr}$, $D_{zz}$, $D_{rz}$ and $D_{zr}$ are functions of
$\vec{x} = (r,z)$.  For details of the implementation of this
anisotropy in the GALPROP code, see Appendix \ref{app:modgalprop}.

All that remains is to relate the diffusion tensor coefficients
$D_{rr}$, $D_{zz}$ and $D_{rz}=D_{zr}$ to the magnetic field model.
\cite{1965P&SS...13....9P} describes the propagation of particles
along ordered field lines in the presence of an irregular component,
and in this case, the diffusion tensor can be written,
\beq 
  D_{ij} = D_0\left(\frac{\nu^2\delta_{ij} + \Omega_i\Omega_j}{\nu^2 +
    \Omega^2}\right),
\eeq
where $D_0$ is the diffusion constant for the isotropic case,
$\delta_{ij}$ is the delta function, $\Omega_{i}$ is the cyclotron
frequency due to the field pointed along the $i$-direction ($\Omega_i
\propto B_i$ and $\Omega^2 = \Omega_i^2 + \Omega_j^2$), and $\nu$ is
the characteristic frequency of deflections by the irregular component
($\nu \propto B_{\rm irr}$).  In our case, we assume for simplicity
that the ordered field is oriented perpendicular to the Galactic
plane, $B_r=0$ and $B_z=B_{\rm ord}$, so that $D_{rz}=D_{zr}=0$.  In
this case, the diffusion tensor becomes,
\beq
  D_{ij} = D_0 \times \left(\begin{array}{cc}
    (1+B_{\rm rat}^2)^{-1} & 0 \\
    0 & 1
  \end{array}\right),
\label{eq:difftensor}
\eeq
where $B_{\rm rat}$ is the ratio of the ordered to irregular field and
we have used the fact that $\Omega/\nu \propto B_{\rm ord}/B_{\rm
  irr}$.  Note that, in the limit of $B_{\rm ord} \rightarrow 0$,
$D_{rr}=D_{zz}=D_0$, and in the limit of $B_{\rm irr} \rightarrow 0$,
$D_{rr} \rightarrow 0$ as desired.  The form of this diffusion tensor
implies that adding an ordered field suppresses diffusion
perpendicular to that field.

For the diffusion tensor \emph{coefficient}, we assume $D_0 \propto
E^{-0.43}$.  However, in contrast to most studies involving GALPROP,
we incorporate the dependence of $D_0$ on $B_{\rm tot}$ as well.  In
particular following \cite{Strong:2007nh},
\beq
  D_0 \propto \left(\frac{B_{\rm irr}}{B_{\rm tot}}\right)^{-2} \times
  r_{\rm gy} = \frac{B_{\rm tot}}{B_{\rm irr}^2},
\eeq
and because $B$ depends on position, $D_0=D_0(r,z)$.  We set the
normalization to be the locally measured value at roughly the locally
measured magnetic field amplitude if the field were completely
irregular, so that our final diffusion coefficient can be written as,
\beq
  D_0 = 2.0\times10^{28}\mbox{ cm$^2$/s} \left(\frac{5\mbox{
      $\mu$G}}{B_{\rm irr}^2/B_{\rm tot}}\right)
  \left(\frac{E}{4.0\mbox{ GeV}}\right)^{-0.5},
\label{eq:diffcoeff}
\eeq
where the normalization is fixed by fitting to the local CR
measurements.

Taken together, Equations \ref{eq:difftensor} and \ref{eq:diffcoeff}
completely define our anisotropic diffusion model and reduce to the
isotropic case when $B_{\rm ord} \rightarrow 0$ and $B_{\rm irr}
\rightarrow$ constant.  For more details about the dependence of
diffusion on the magnetic field, see Appendix \ref{app:diff_dep_B}.

Lastly we note that, in all of our models, we use a box height $L_{\rm
  box} = \pm20$ kpc.  This is not directly comparable to the usual box
heights ($\sim 4$ kpc) discussed in the literature, because the ``free
escape'' of electrons outside the Galactic disk is taken into account
by the spatial dependence of the diffusion tensor.  This is in
agreement with findings by the \emph{Fermi} team regarding diffuse IC
away from the GC \citep{PorterTeVPA} and is in fact a more appropriate
box size.  This also alleviates the problem of ``squashed''
morphologies that are typical of smaller box heights when the CR
density at the boundary is set to zero.

%%%%%%%%%%%%%%%%%%%%%%%%%%%%%%%%%%%%%%
%%  Dark matter annihilation model  %%
%%%%%%%%%%%%%%%%%%%%%%%%%%%%%%%%%%%%%%

\subsection{DM annihilation model}

In Equation \ref{eq:Propagation_equation}, the source term $Q$ is the
rate of $e^+e^-$ injection by DM annihilations and is given by
\beq
  Q(r,z) = \frac{1}{2} \sigmav \frac{dN}{dE}
  \left(\frac{\rho(r,z)}{M_{\chi}}\right)^2,
\label{eq:injection}
\eeq
where $dN/dE$ is the injection spectrum and $\rho$ is the Galactic DM
halo.  We assume a prolate Einasto \citep{Einasto} halo,
\beq
  \rho(r,z)\propto
  \exp\left[-\frac{2}{\alpha}\left(\left(\frac{r^{2}}{r_{c}^{2}} +
    \frac{z^{2}}{z_{c}^{2}}\right)^{\alpha/2} -
    \frac{R_{\odot}^{\alpha}}{r_{c}^{\alpha}}\right)\right],
\label{eq:DeformedEinasto}
\eeq
with $z_{c}/r_{c}=2.0$, $z_{c}=27$ kpc, and $\alpha=0.17$
\citep{Merritt:2005xc}.  The overall normalization is set so that the
local DM density is $\rho(R_{\odot},0) = 0.4$ GeV/cm$^3$
\citep{Catena:2009mf}.

The injection spectrum $dN/dE$ is governed by the specific particle
model.  In our case, we use XDM \citep{Finkbeiner:2007kk} as our
fiducial model, with $M_{\chi}=1.2$ TeV, an annihilation channel
$\chi\chi \rightarrow \phi\phi$, $\phi \rightarrow e^{+}e^{-}$, and
with branching ratio 1 \citep[hereafter, XDM $e^{\pm}$;
  see][]{Cholis:2008vb, Cholis:2008qq}.  In this model, $\phi$ is a
vector boson with $m_{\phi} \le 2m_{\mu}$ that is the force carrier
responsible for the velocity dependent Sommerfeld enhancement
\citep{ArkaniHamed:2008qn,Pospelov:2008jd}.  We do not include
specific dynamics for the host halo, but we do assume that the
velocity dispersion (and hence the Sommerfeld enhancement or ``boost
factor'') as well as substructure contribution is flat with radius.
We define this boost factor $BF$ as,
\beq
  BF = \frac{\sigmav}{3 \times 10^{-26}\mbox{ cm$^3$/s}}.
\label{eq:boostfactor}
\eeq
This model for the DM particle has $E^2 dN/dE \propto E^2$ as required
by the CR, microwave, and gamma-ray data \citep{2010PhRvD..82b3518L}.

%%%%%%%%%%%%%%%%%%%%%%%%%
%%  Fitting Procedure  %%
%%%%%%%%%%%%%%%%%%%%%%%%%

\subsection{Fitting Procedure}

We follow a similar procedure as that outlined in Appendix C of
\cite{dobler:2010fh}.  Specifically, we generate a synthetic sky map,
\beq 
  S(E) = A_{\rm loE} \times E_{0.5}^{1.0} + A_{\rm gp} \times G(E) +
  U(E),
\eeq
where $A_{\rm loE}$ and $A_{\rm gp}$ are the amplitudes of the
\emph{Fermi} 0.5-1.0 GeV map and the GALPROP map at mean energy $E =
\sqrt{E_0 E_1}$ respectively and $U(E)$ is a uniform background, and
convert to a synthetic counts map $\mu(E) = S(E) \times ({\rm mask})
\times ({\rm exposure})$.  We then minimize the log-likelihood,
\beq
  \ln {\mathcal L} = \sum_i [k_i \ln\mu_i - \mu_i - \ln(k_i!)],
\eeq
where $k_i$ is the map of observed counts at pixel $i$, over the
parameters $A_{\rm loE}$ and $A_{\rm gp}$.  When comparing maps at
different energies, it is important to smooth the templates and data
to a common beam full-width half-maximum (FWHM).  All of our maps use
1.6 years of data, are smoothed to 2 degrees, and for the
$E_{0.5}^{1.0}$ map, we use only ``front'' converting events
\citep[see][]{dobler:2010fh}.

%%%%%%%%%%%%%%%
%%  Results  %%
%%%%%%%%%%%%%%%

\section{Results}
\label{sec:results}

\bpm
  \centerline{
    \includegraphics[width=0.49\textwidth]{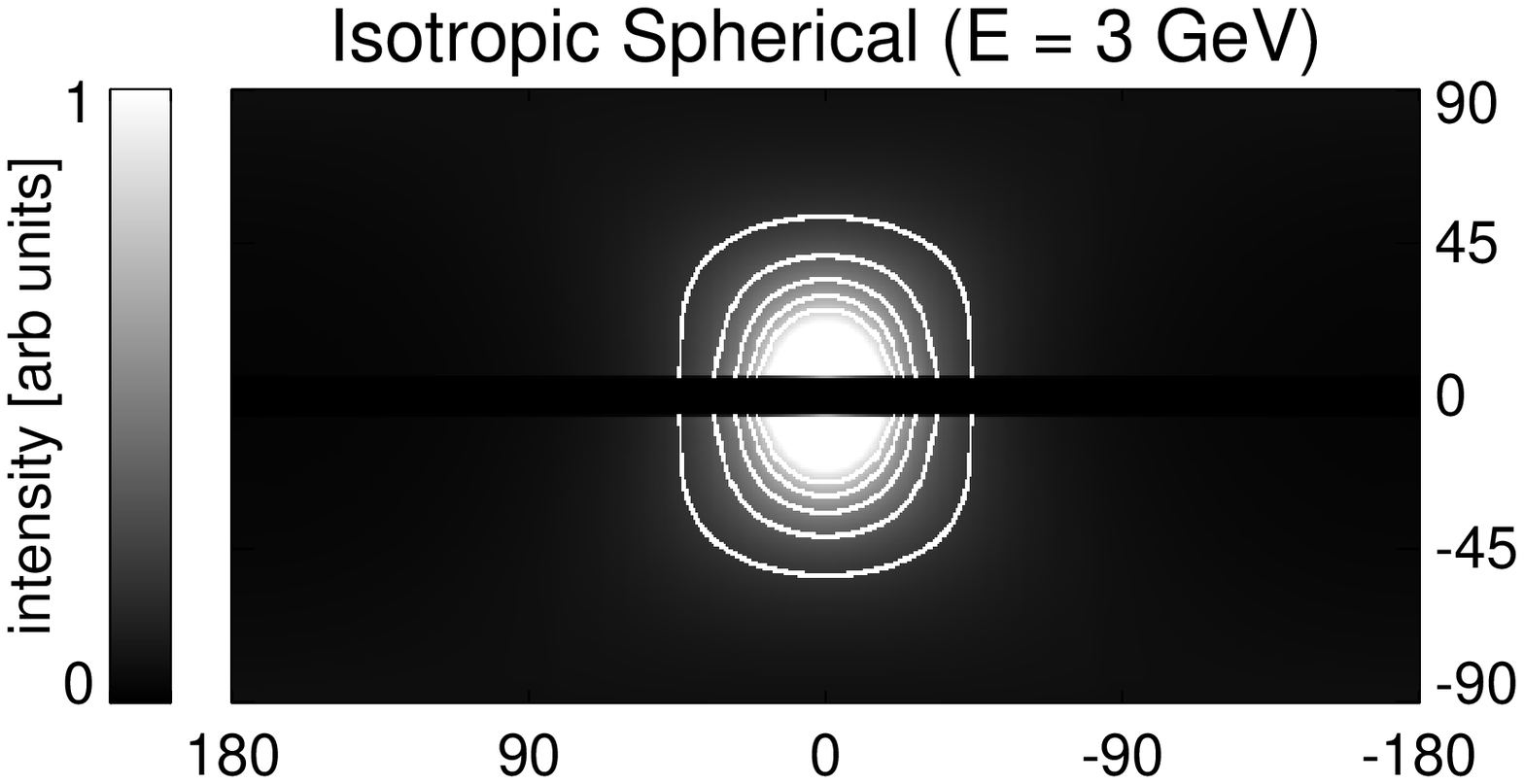}
    \includegraphics[width=0.49\textwidth]{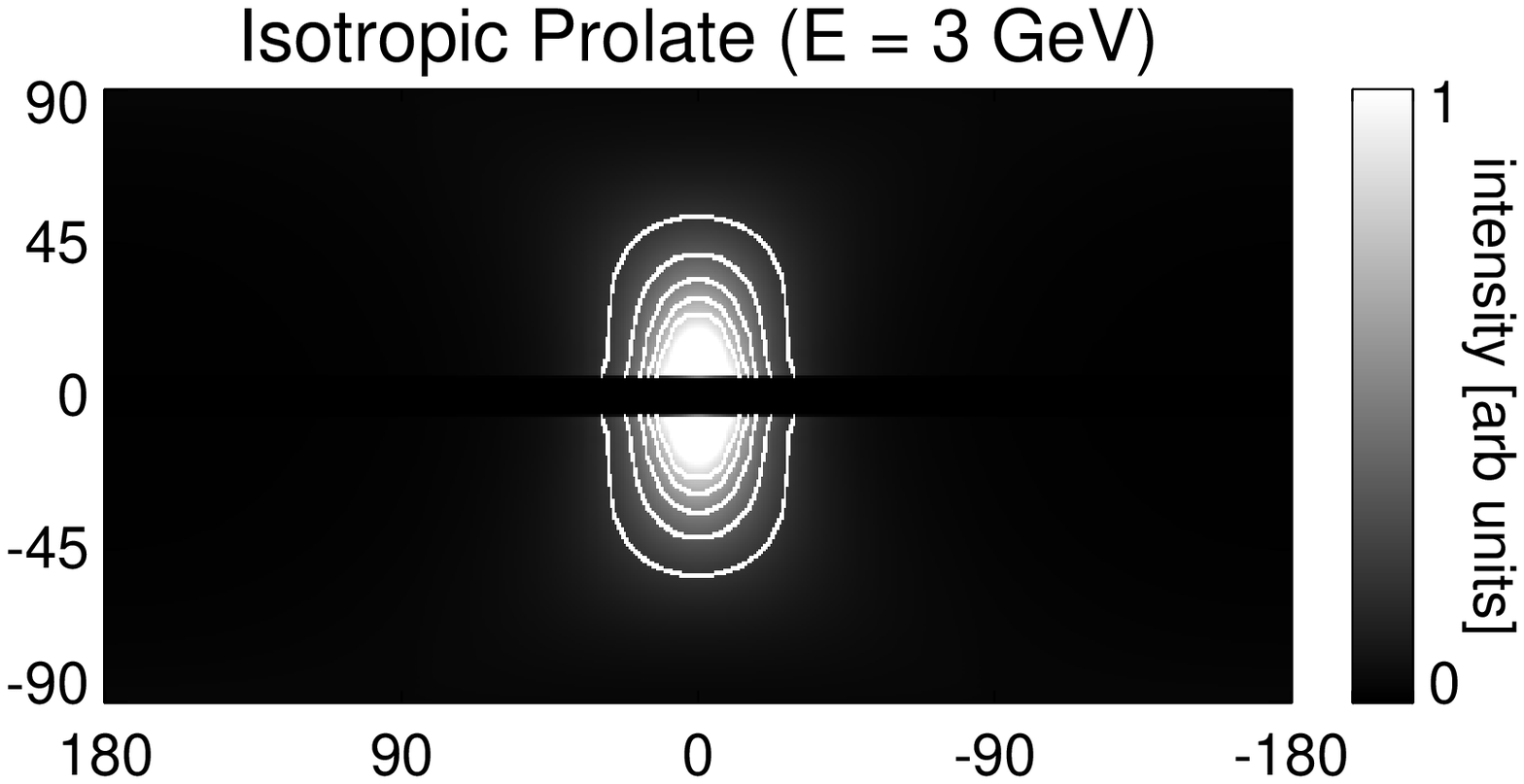}
  }
  \centerline{
    \includegraphics[width=0.49\textwidth]{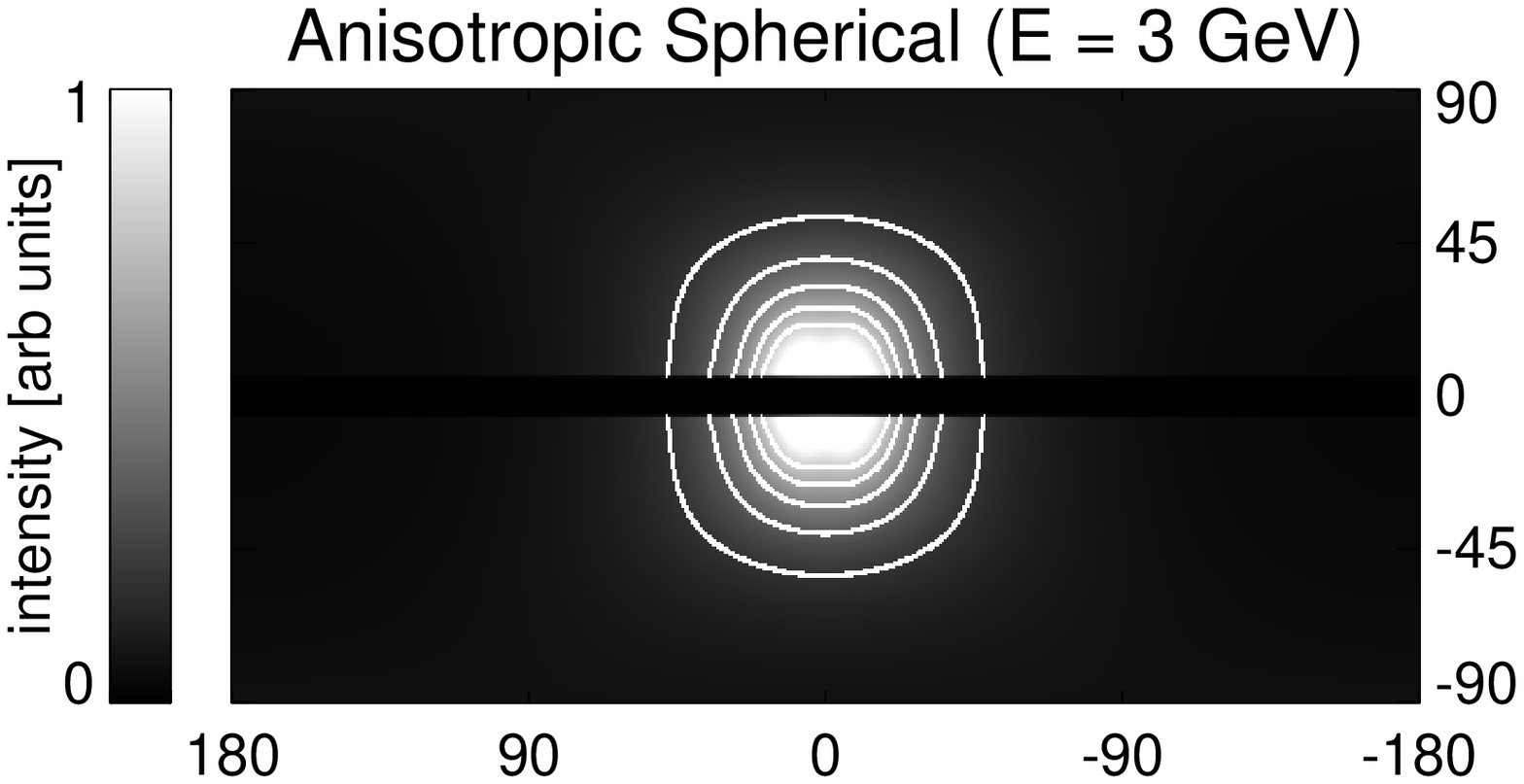}
    \includegraphics[width=0.49\textwidth]{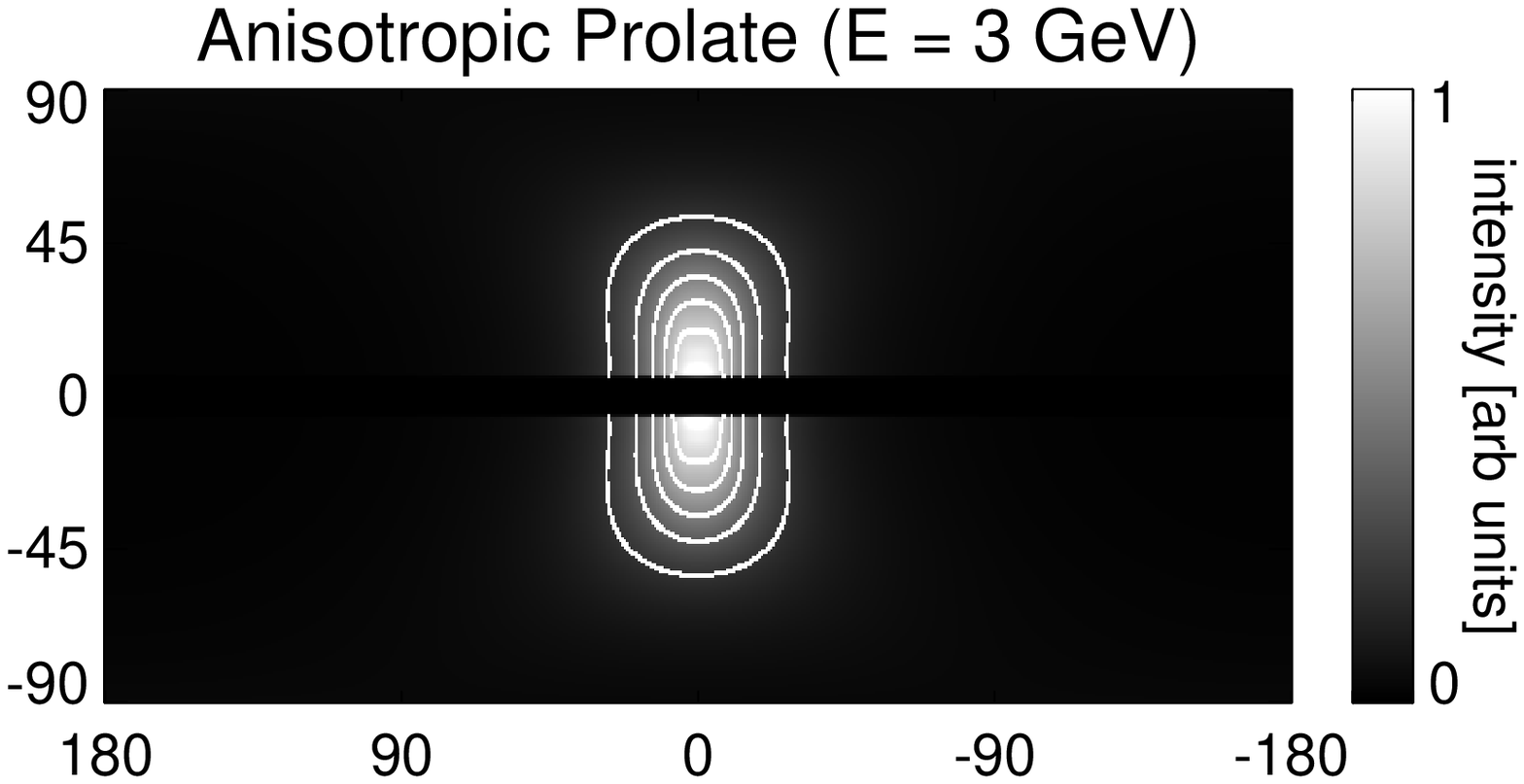}
  }
\caption{
  GALPROP IC at 3 GeV due to $\epp$ production by DM annihilations
  with different assumptions about the halo shape and diffusion model:
  a spherical Einasto halo with isotropic diffusion (\emph{upper
    left}), an axis ratio 2 prolate halo with isotropic diffusion
  (\emph{upper right}), a spherical halo with anisotropic diffusion
  effects (\emph{lower left}), and a prolate halo with anisotropic
  diffusion effects (\emph{lower right}).  All plots are arbitrarily
  normalized to the same intensity at $(\ell,b) = (0,50)$ degrees.
  The spherical halos are clearly inconsistent with the haze
  morphology (see the right hand panels of \reffig{compmorph}) while
  the prolate halos provide a significantly improved fit.  In
  particular, the anisotropic diffusion case gives a morphology that
  has both the observed axis ratio and concentration.
}
\label{fig:addanisotropy}
\epm

\reffig{addanisotropy} shows the GALPROP IC map for $E = 3.0$ GeV and
for various assumptions about the dark halo prolateness and
anisotropic diffusion.  For the case of a spherical halo with
isotropic diffusion (completely tangled magnetic field), the resultant
IC signal is largely spherical.  The same is true for our anisotropic
model with a spherical halo, implying that diffusion effects alone
cannot create the observed morphology.  In fact prolate halos lead to
IC morphologies which very closely resemble the haze morphology.  In
detail, we find that the prolate halo with isotropic diffusion is
overly concentrated towards the center and that the best morphological
match to the data comes from using a prolate halo with anisotropic
diffusion.

\bpm
  \centerline{
    \includegraphics[width=0.49\textwidth]{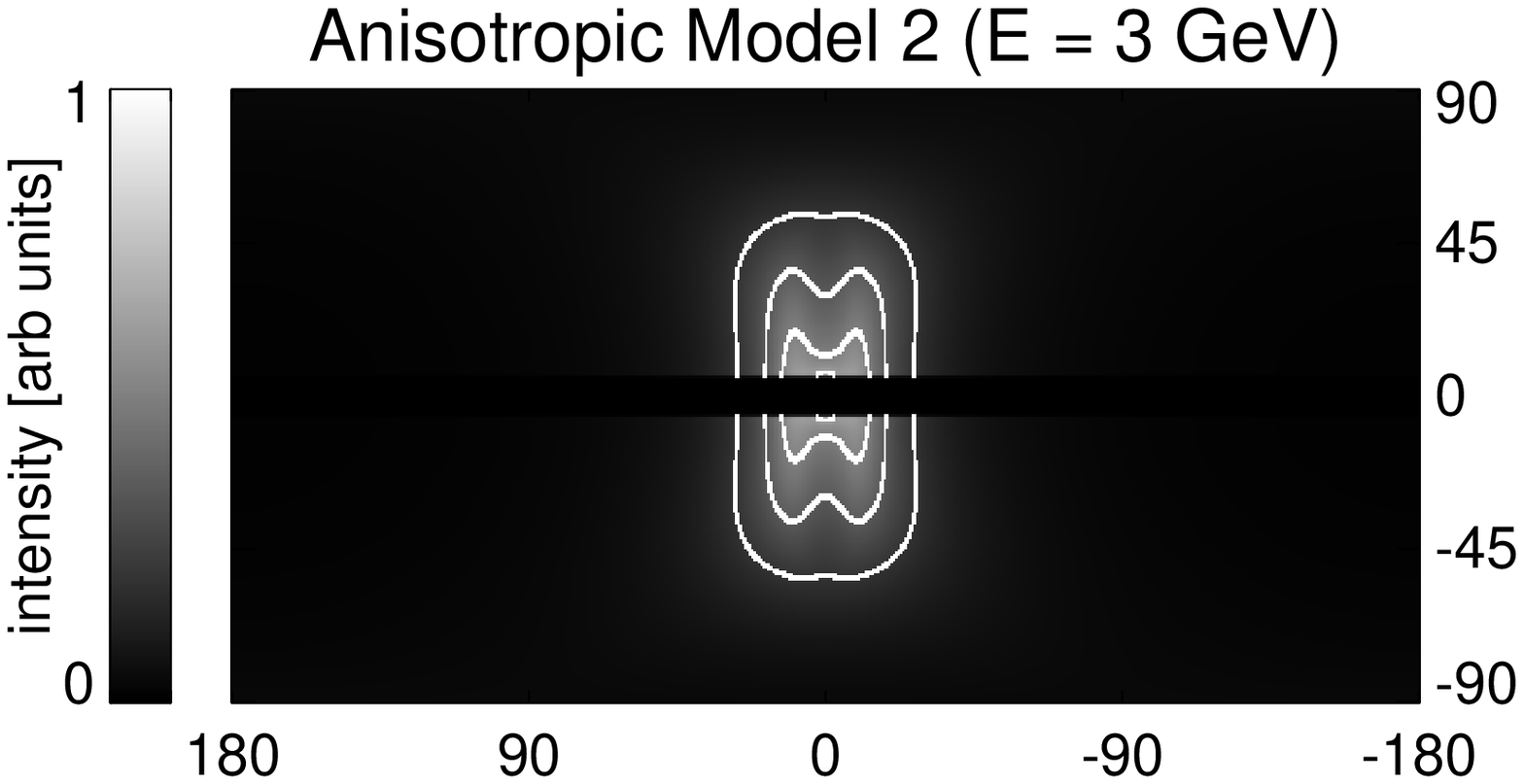}
    \includegraphics[width=0.49\textwidth]{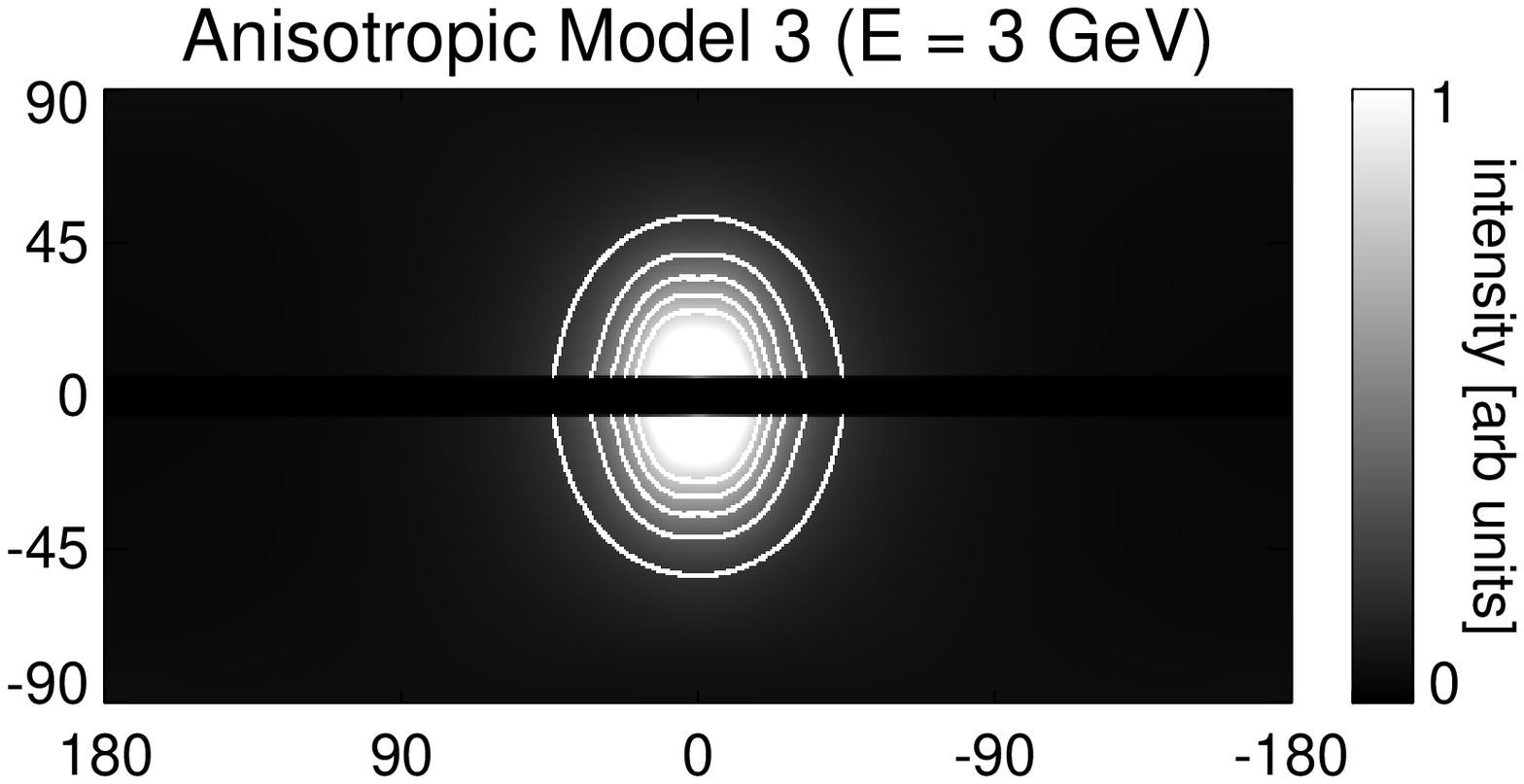}
  }
  \centerline{
    \includegraphics[width=0.49\textwidth]{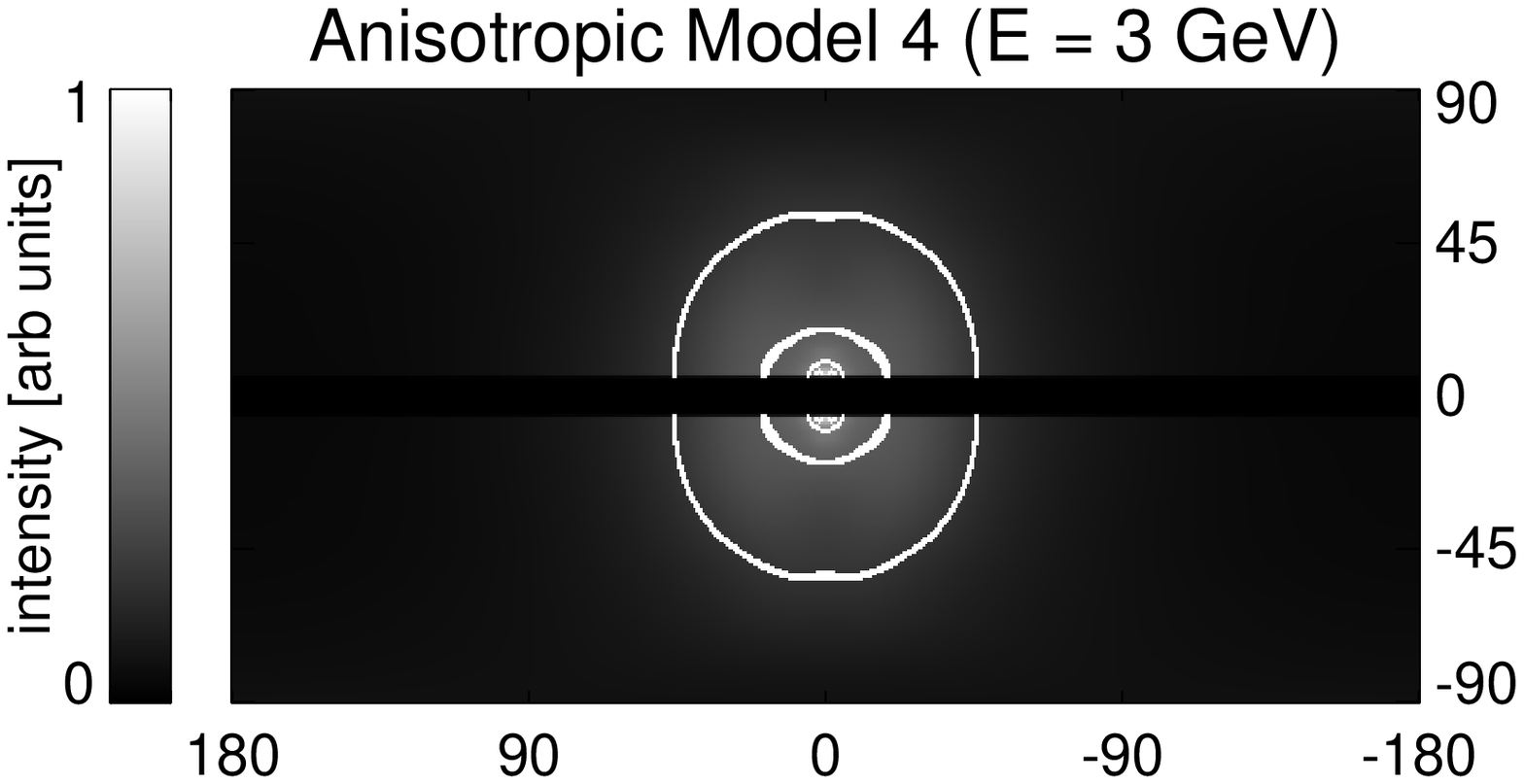}
    \includegraphics[width=0.49\textwidth]{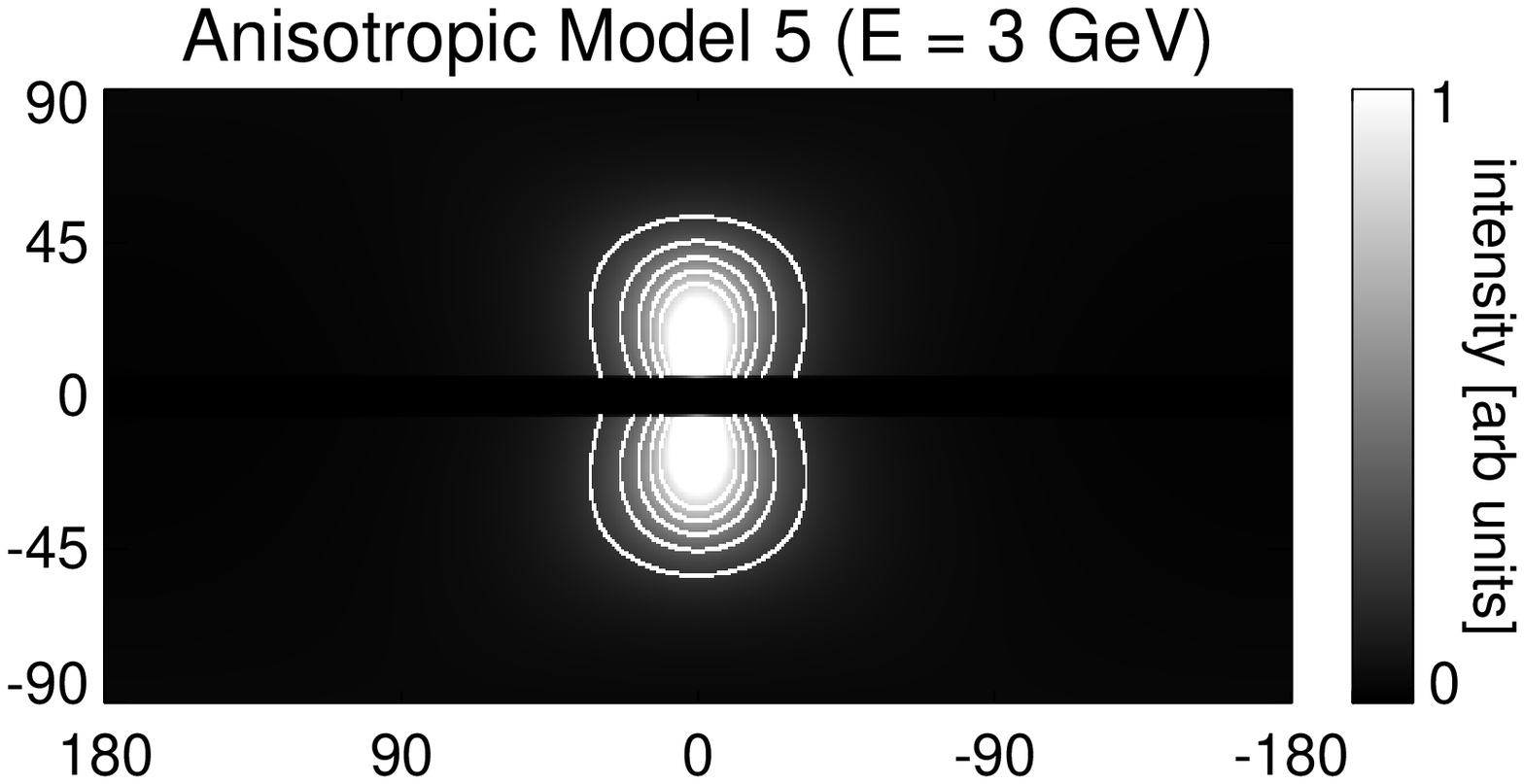}
  }
\caption{
  The same as the bottom right panel of \reffig{addanisotropy} but for
  several different models of the ordered magnetic field (see Table
  \ref{tbl:Bfieldmods}).  Different magnetic field models can lead to
  various IC morphologies including forked (\emph{top left}, due to
  increased synchrotron losses towards $r=0$ kpc), circular and
  centrally concentrated (\emph{top right}), circular and more uniform
  (\emph{lower left}), and also more \emph{hourglass}-shaped
  (\emph{lower right}).  See \refsec{results} for a description.
}
\label{fig:varyBfieldmorph}
\epm

The detailed assumptions on the B-field morphology, and thus on the
spatial dependence of the diffusion, can have a strong effect on the
observed morphology of the IC emission as is shown in
\reffig{varyBfieldmorph} where we present the IC maps at 3 GeV for
four distinctively different $B_{ord}$ assumptions from those of
Equation \ref{eq:bfieldord}.  The specific magnetic field model can
lead to various IC morphologies from more uniform to more centrally
concentrated and from more elliptical to more circular.  In addition,
for fields with a strong ordered component towards $r=0$ kpc, forked
morphologies (due to increased synchrotron losses towards $r=0$ kpc)
are found.  Interestingly, for relatively modest changes to our
magnetic field parameters, we can also reproduce an \emph{hourglass}
shape reminiscent of the ``bubble'' shape in \cite{Su:2010qj}.  Note
that all models use an identical prolate dark matter halo; the
variations in shape are due exclusively to magnetic field effects on
the diffusion and relative energy losses to synchrotron and IC.

\bpm
  \centerline{
    \includegraphics[width=0.49\textwidth]{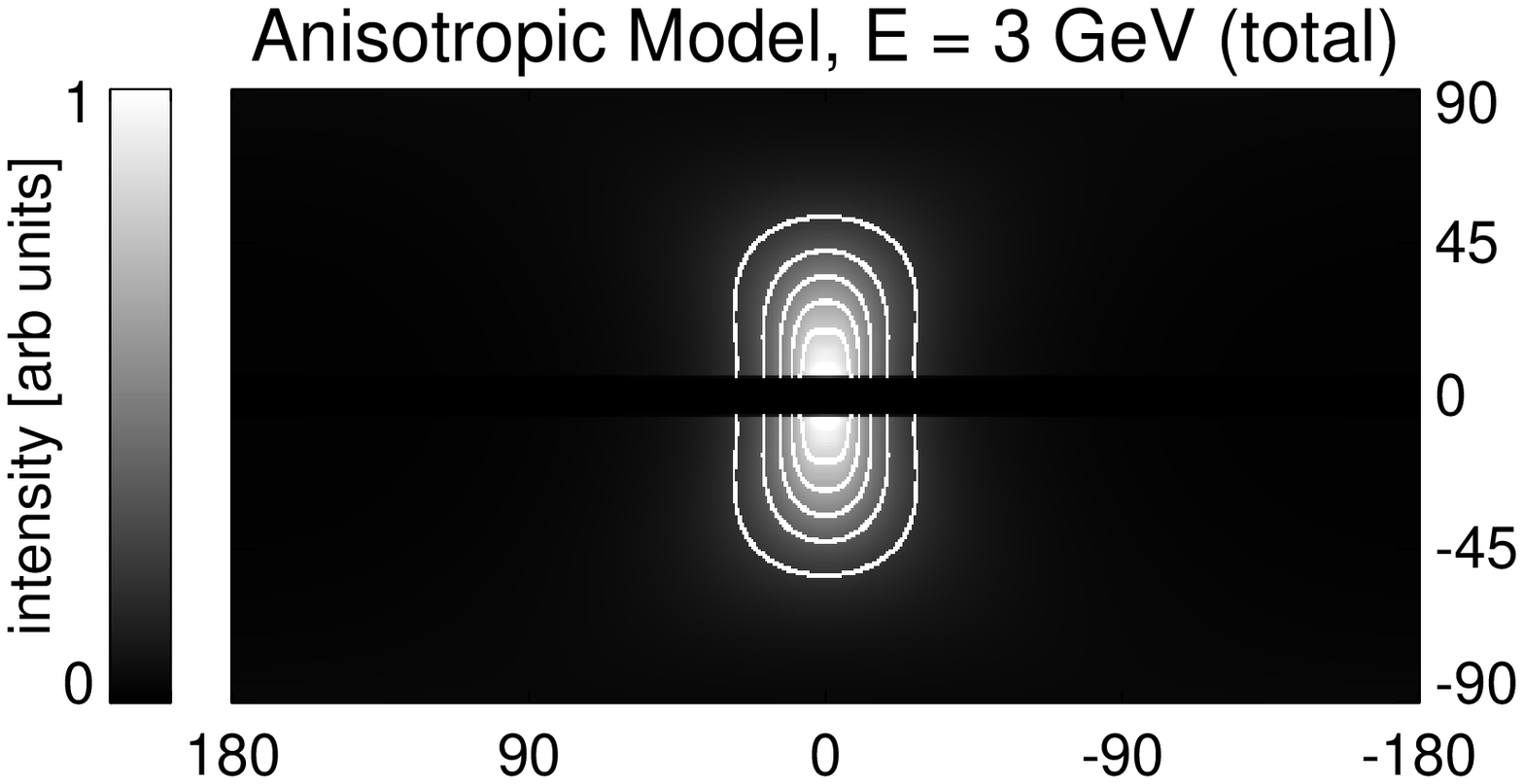}
    \includegraphics[width=0.49\textwidth]{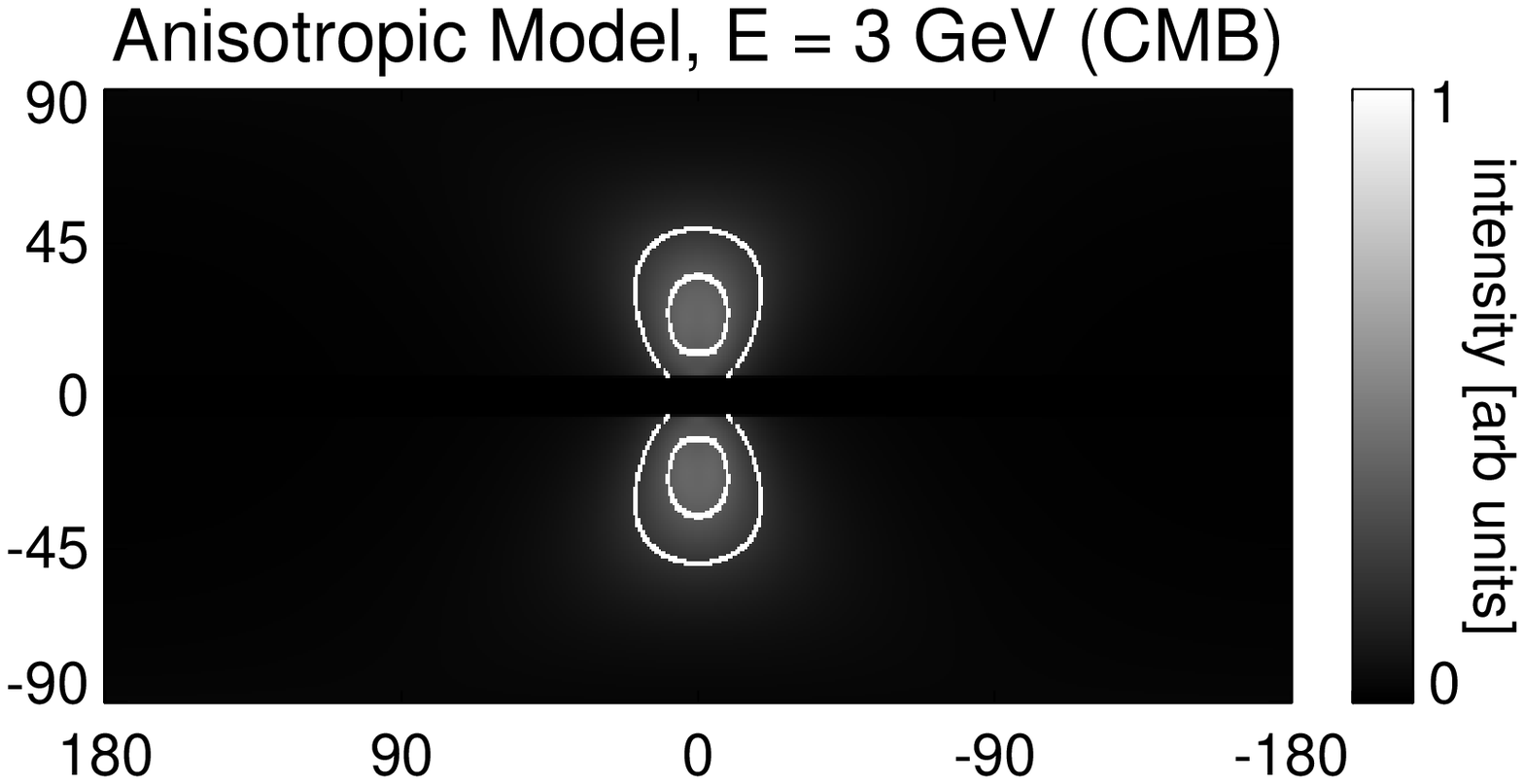}
  }
  \centerline{
    \includegraphics[width=0.49\textwidth]{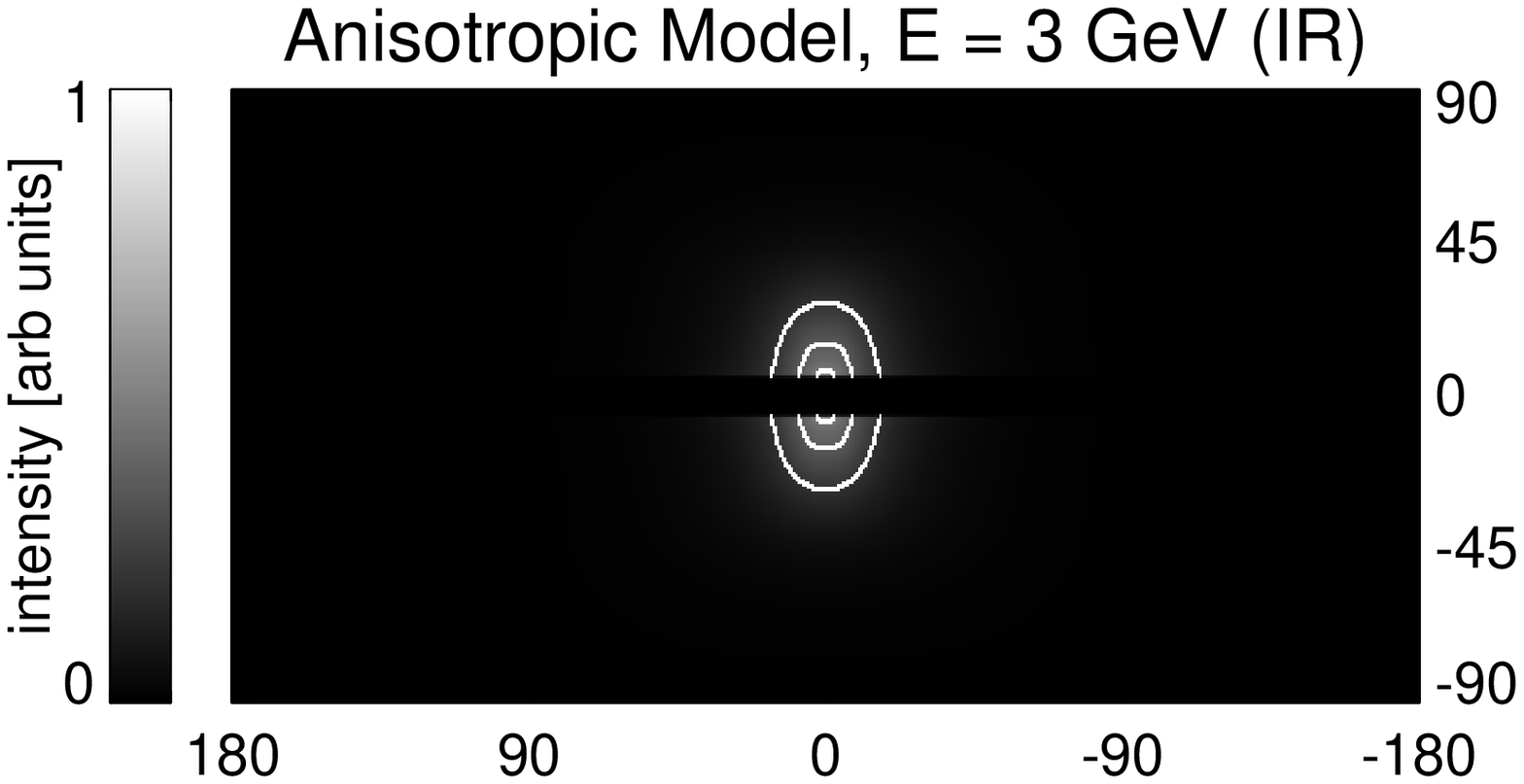}
    \includegraphics[width=0.49\textwidth]{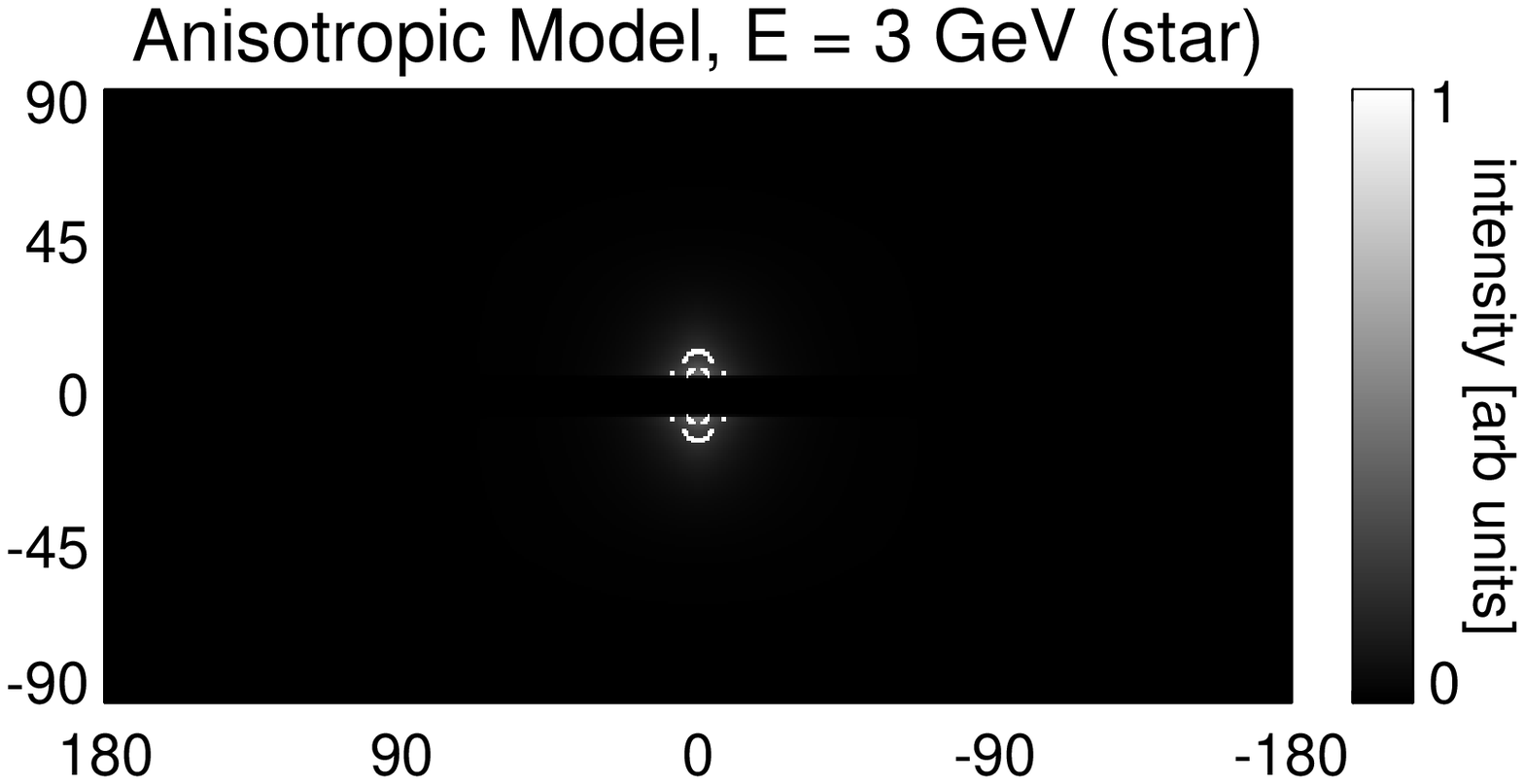}
  }
\caption{
  The full anisotropic, prolate model at 3 GeV (\emph{top left}) as
  well its individual components broken down into photons from IC
  scattered CMB (\emph{top right}), IR (\emph{bottom left}), and
  starlight (\emph{lower right}) photons.  All maps use the same
  stretch, normalization, and contour intervals.  The CMB component in
  particular has a distinctly ``bubble''-like morphology.  Thus, when
  using template regression techniques to assess the underlying
  morphology of the haze, care must be taken not to regress out emission
  from IR and starlight components while leaving only the CMB
  component.
}
\label{fig:iccomponents}
\epm

This IC emission is the combination of electrons scattering CMB, IR,
and starlight photons.  Each of these ISRF components has a distinct
morphology, and so the IC emission from each will also have a
different morphology.  In fact, since the starlight and IR photons are
mostly confined to the plane, the high latitude IC emission is due
primarily to scattering of CMB photons.  This is borne out in
\reffig{iccomponents} which shows the morphology of each of the IC
components.  The starlight and IR IC photons are concentrated much
more towards the GC while the CMB IC photons extend to much higher
latitudes.  Furthermore, it is interesting to note the distinct
``bubble''-like morphology of the CMB component.  The implication here
is that template fits like those used in \cite{dobler:2010fh} and
\cite{Su:2010qj}, which use external templates that are concentrated
towards the GC could potentially absorb the starlight and IR
morphologies, while leaving the CMB morphology which appears more
bubble-like.  That is, if the intrinsic haze morphology is more
oval-shaped, pulling out only the CMB component would may leave a
bubble morphology.

\bpm
  \centerline{
    \includegraphics[width=0.49\textwidth]{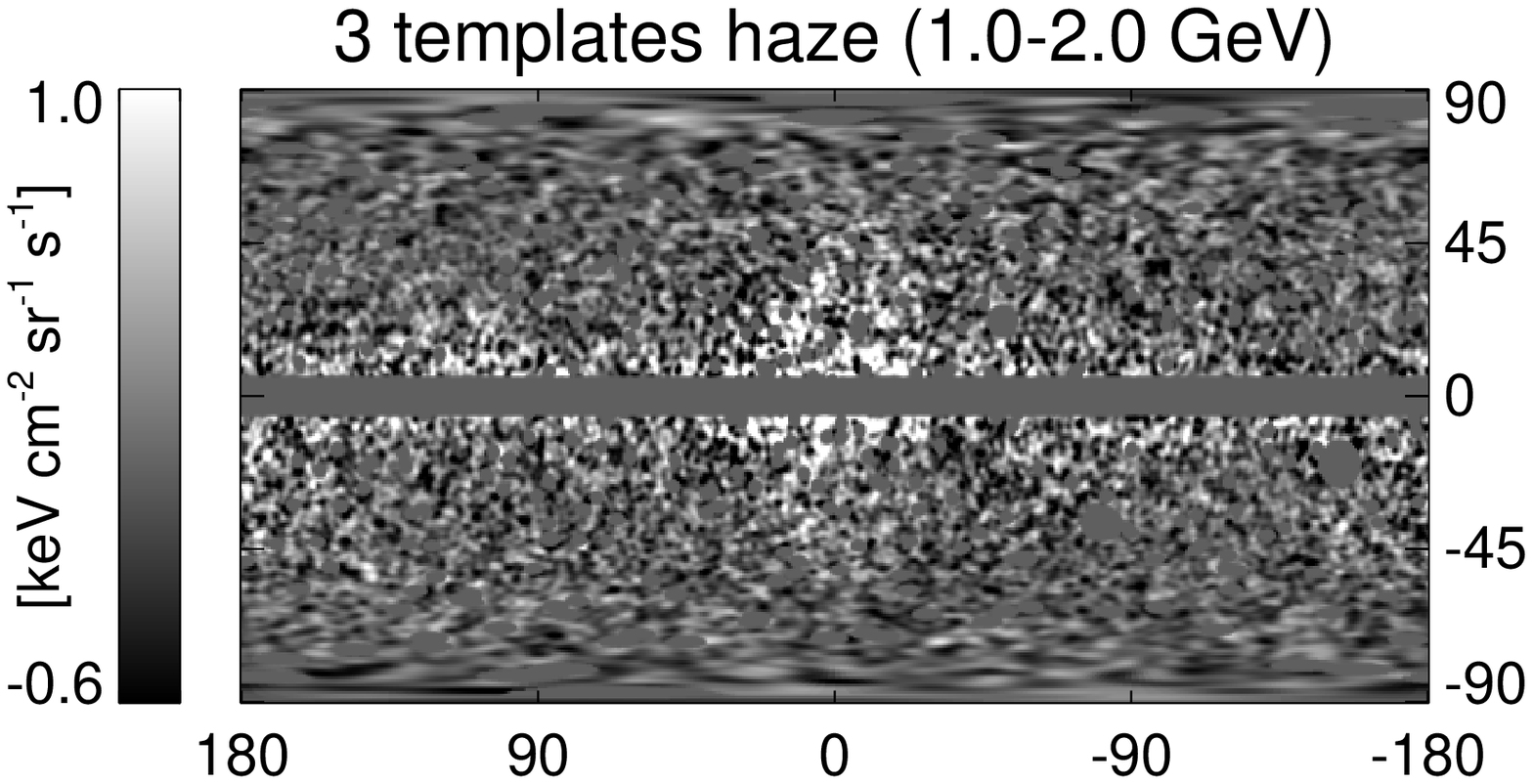}
    \includegraphics[width=0.49\textwidth]{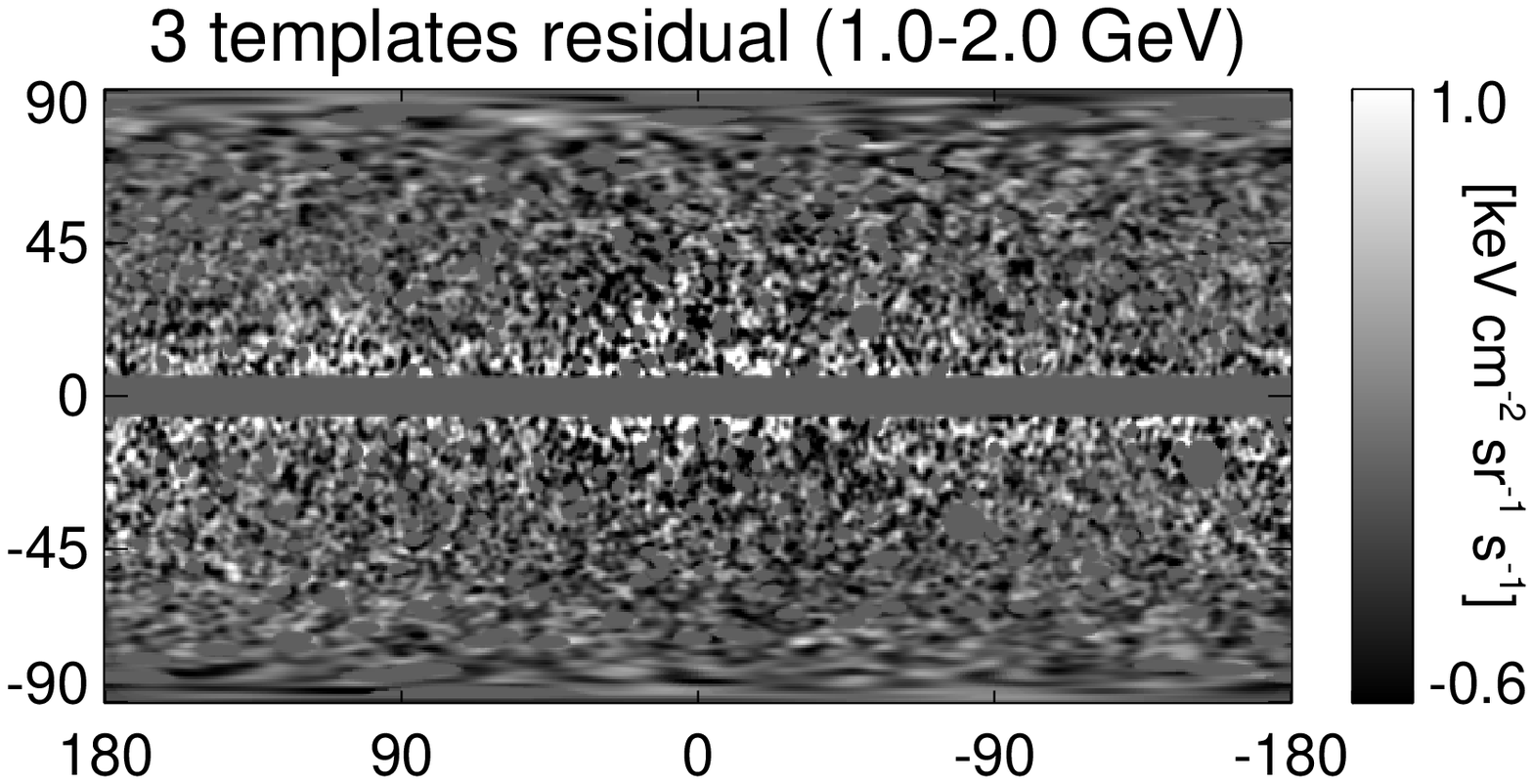}
  }
  \centerline{
    \includegraphics[width=0.49\textwidth]{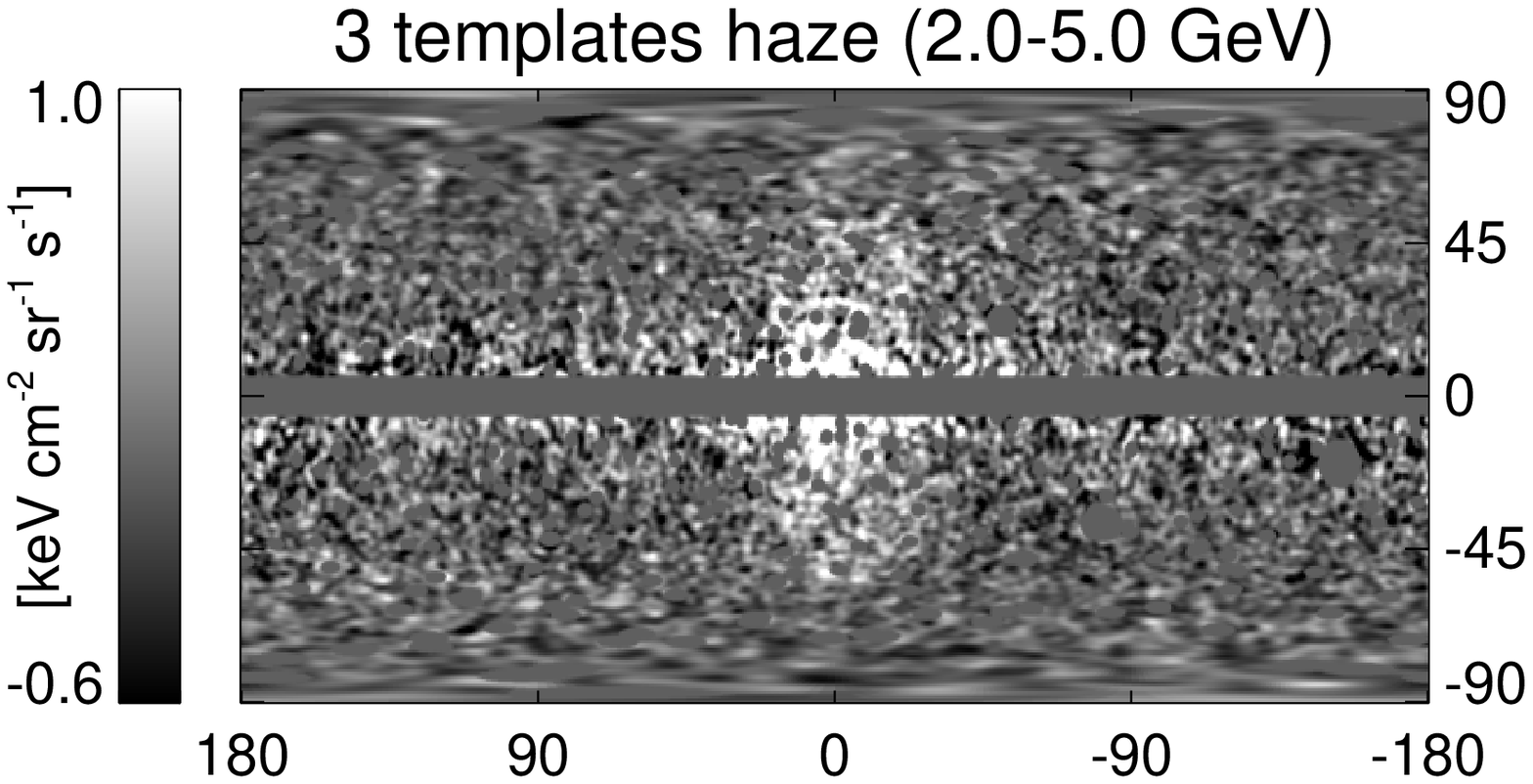}
    \includegraphics[width=0.49\textwidth]{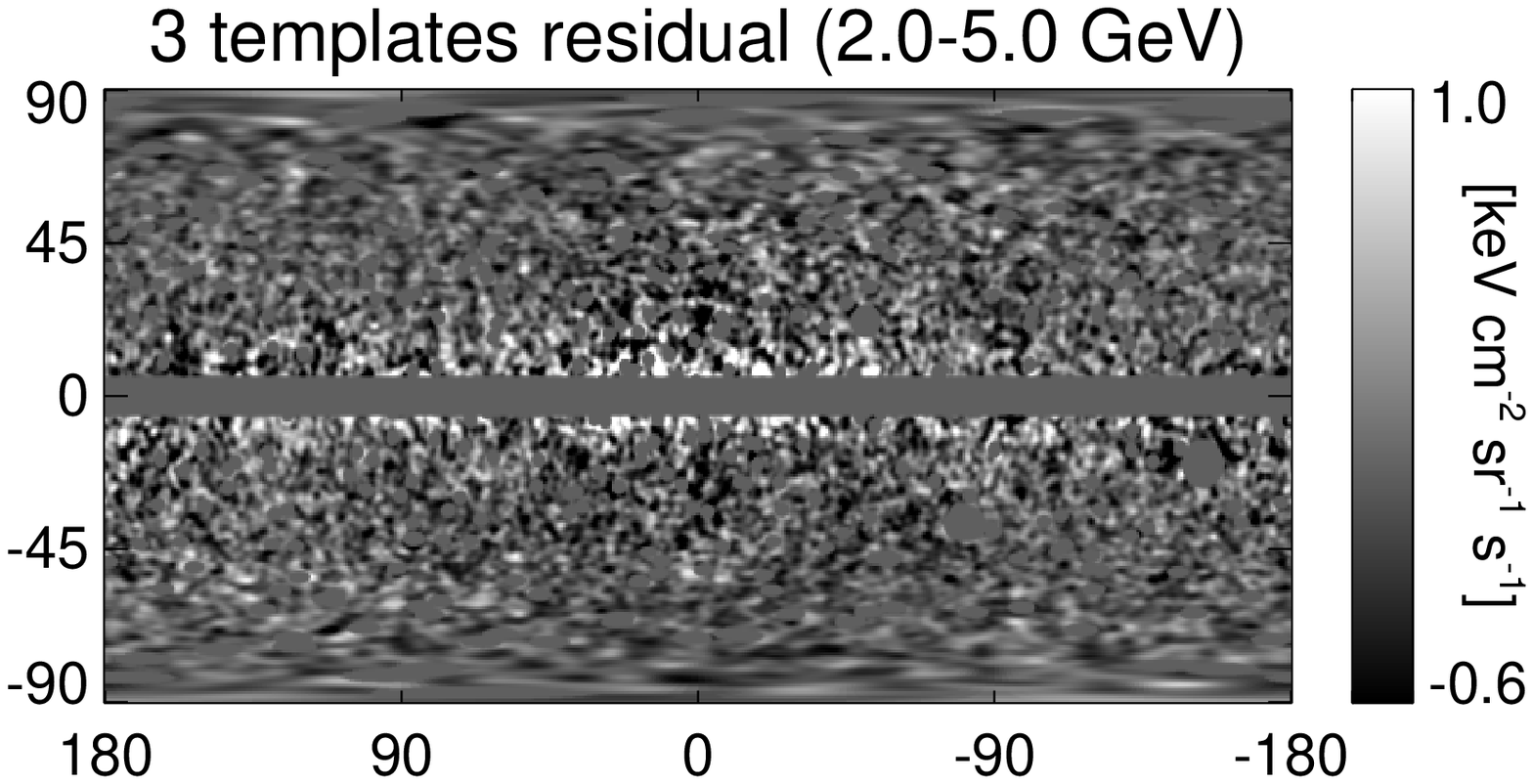}
  }
  \centerline{
    \includegraphics[width=0.49\textwidth]{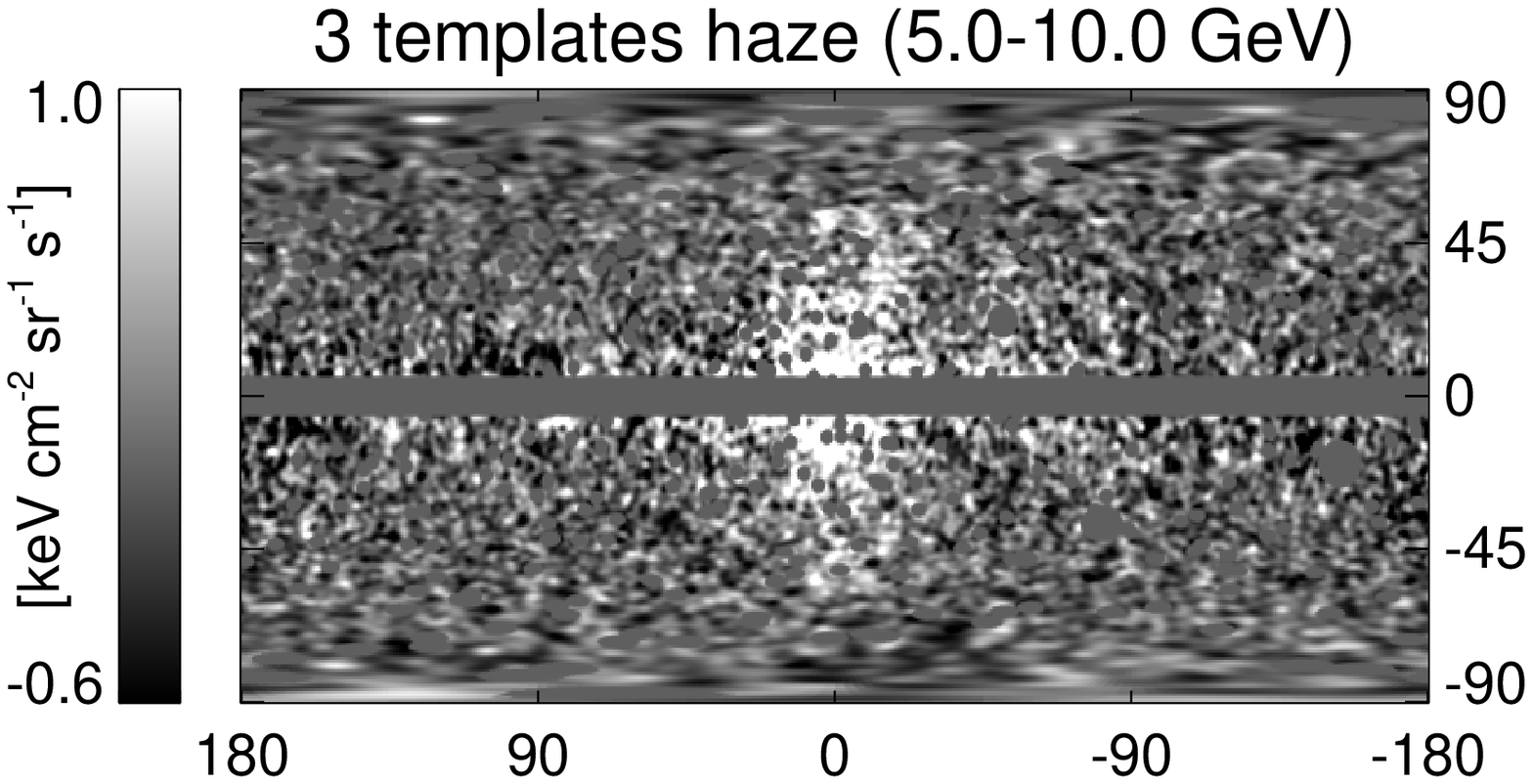}
    \includegraphics[width=0.49\textwidth]{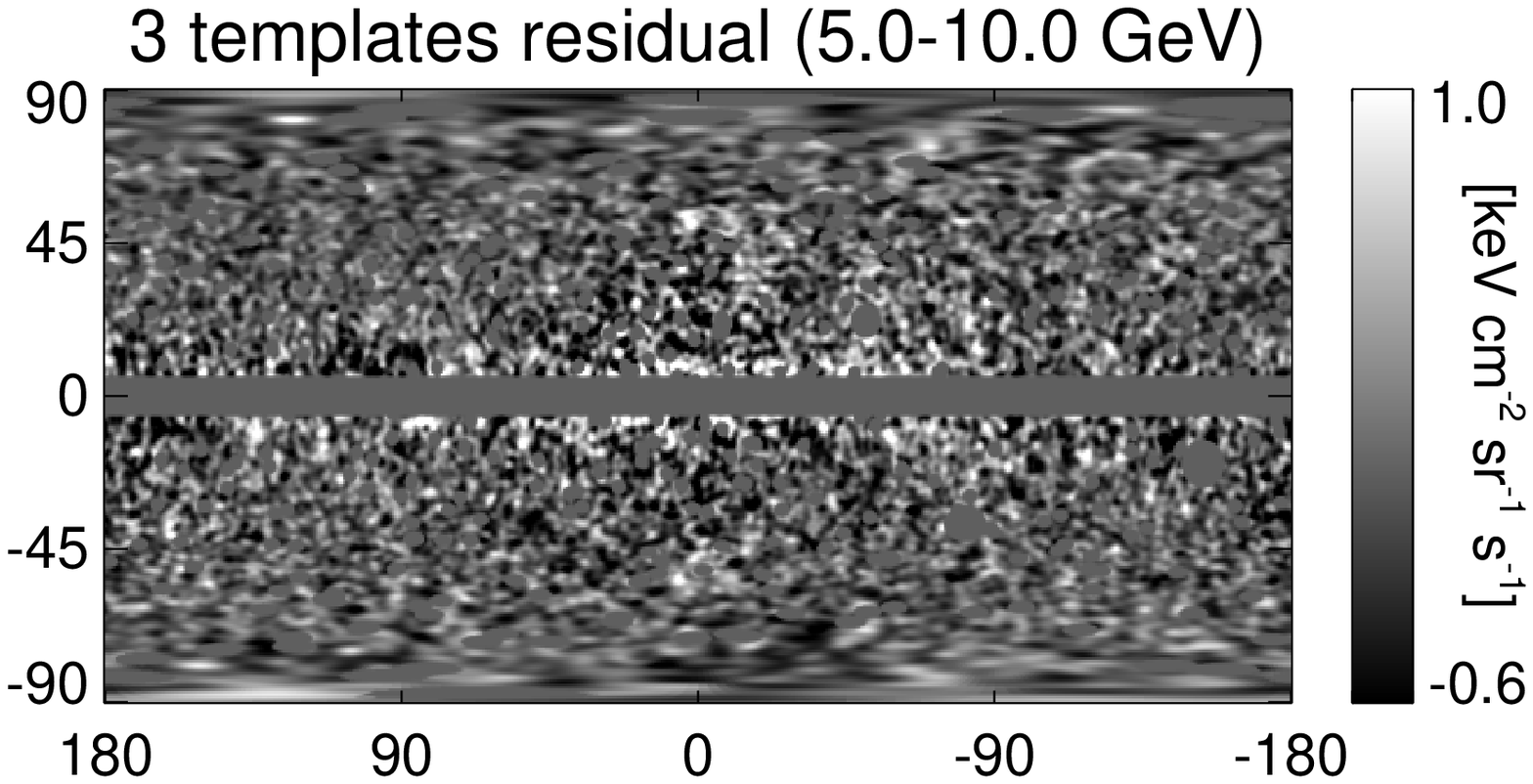}
  }
  \centerline{
    \includegraphics[width=0.49\textwidth]{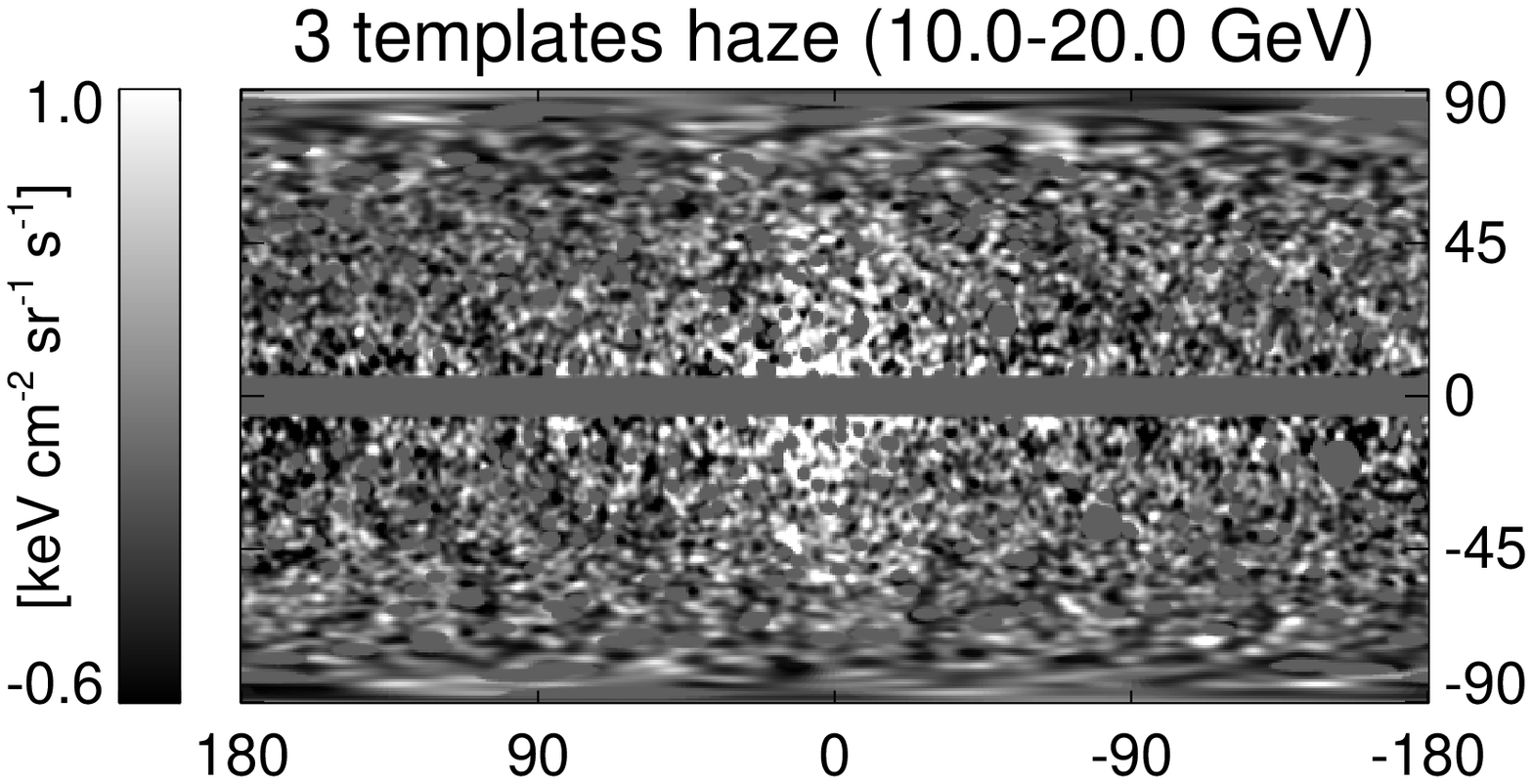}
    \includegraphics[width=0.49\textwidth]{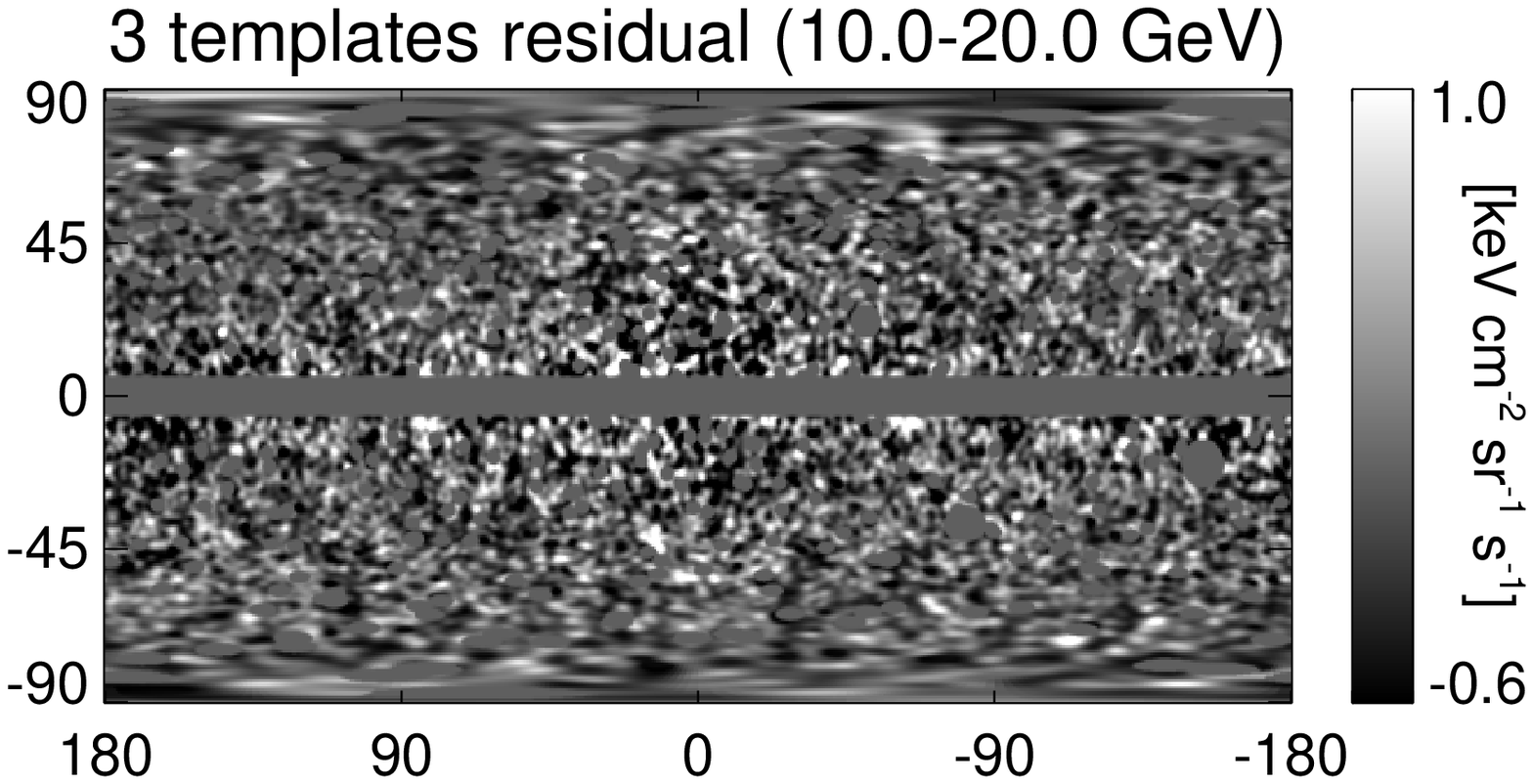}
  }
\caption{
  The haze (\emph{left column}) and residuals (\emph{right column}) of
  our three template fit in energy bins from 1.0 to 20.0 GeV.  The
  haze maps clearly show the strong haze residual with an axis ratio
  $\approx$2 that is in reasonably good morphological agreement with
  our anisotropic dark matter model (see \reffig{iccomponents} top
  left panel).  This is borne out in the residual maps which show a
  residual consistent with noise at latitudes above 20$^o$.  Close
  inspection reveals a slight over-subtraction towards the center due
  to the fact that our model does not explicitly include an ``edge''
  at $b\approx\pm50^o$ as is seen in the data.
}
\label{fig:residuals}
\epm

In \reffig{residuals} we show the residual ``haze'' map,
\beq
  {\mathcal H}_{E_0}^{E_1} = E_{E_0}^{E_1} - S_{E_0}^{E_1} + A_{\rm
    gp} \times G(E),
\eeq
as well as the residual map,
\beq
  {\mathcal R}_{E_0}^{E_1} = E_{E_0}^{E_1} - S_{E_0}^{E_1}.
\eeq
As shown in the figure, the three component model provides a
remarkably good fit to the data.  There is some residual
over-subtraction due to the fact that the \emph{Fermi} haze appears to
have an ``edge'' at roughly $|b|\sim50$ deg.  This feature cannot be
reproduced exactly by our models which tend to be slightly more
diffuse.  This lack of an edge pushes the fit to slightly over
subtract the GALPROP haze contribution.  Indeed, astrophysical models
such as winds \citep{crocker:2011gw} or jets would also either not
have an edge or, in the case of jets, likely have a shock heated edge
with a harder spectrum which is not clearly seen in the data
\citep{Su:2010qj}.  Despite this, our fit removes 96\%, 89\%, and 69\%
of the variance over pixels with $|b|>5^o$at $E = $ 2-5, 5-10, and
10-20 GeV respectively (see \reffig{residuals}).

\bpm
  \centerline{
    \includegraphics[width=0.49\textwidth]{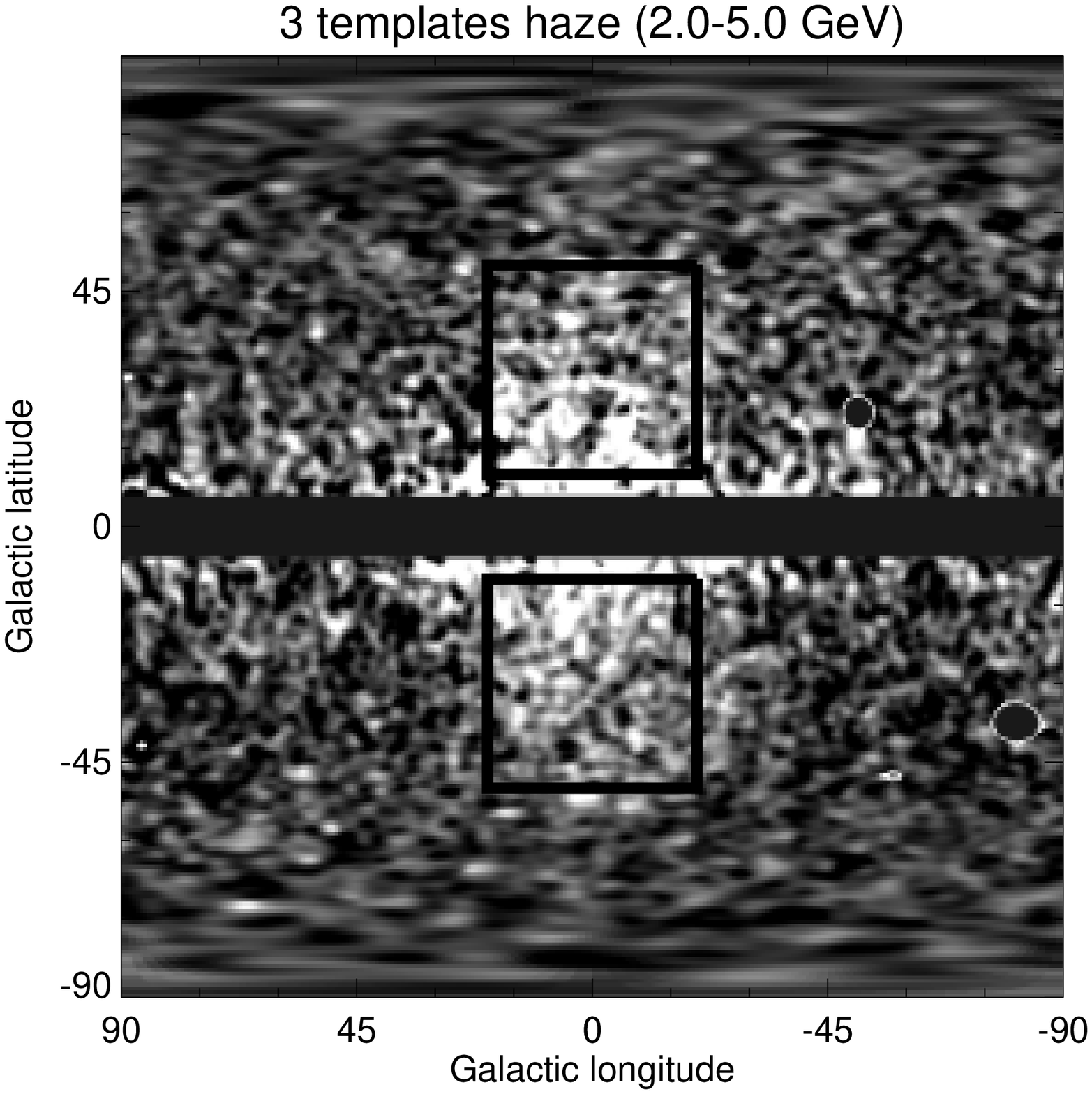}
    \includegraphics[width=0.49\textwidth]{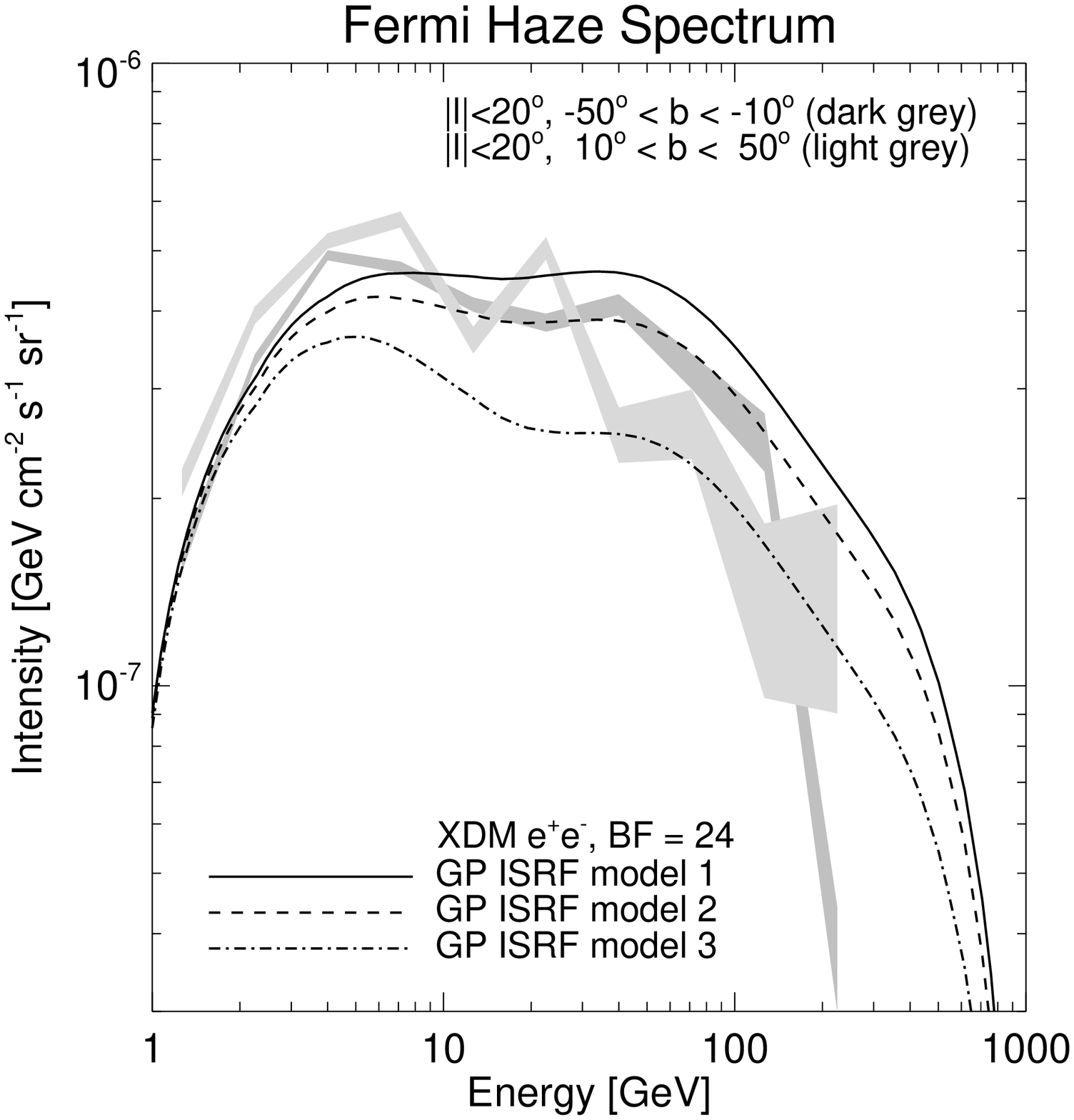}
  }
  \centerline{
    \includegraphics[width=0.49\textwidth]{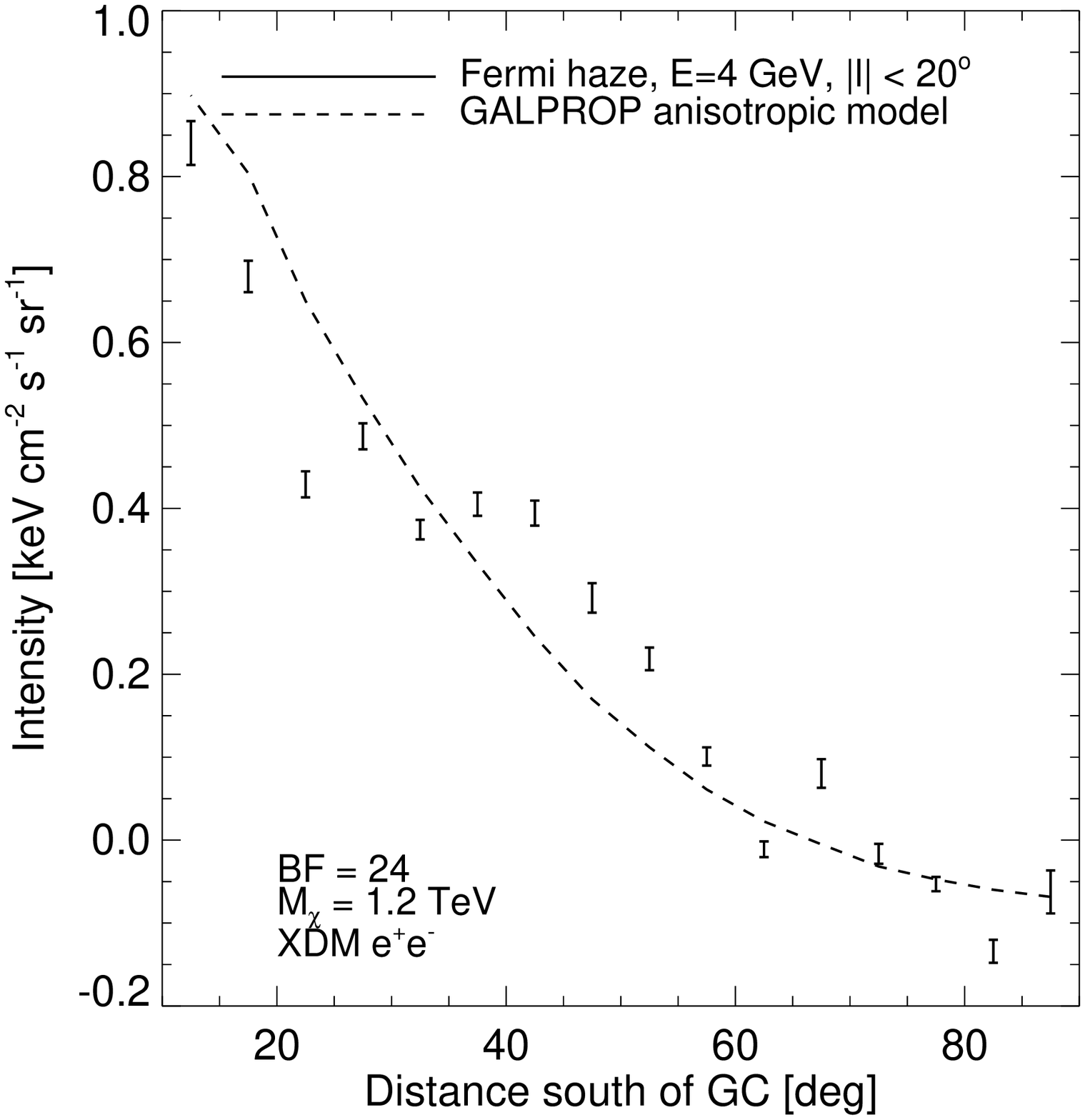}
    \includegraphics[width=0.49\textwidth]{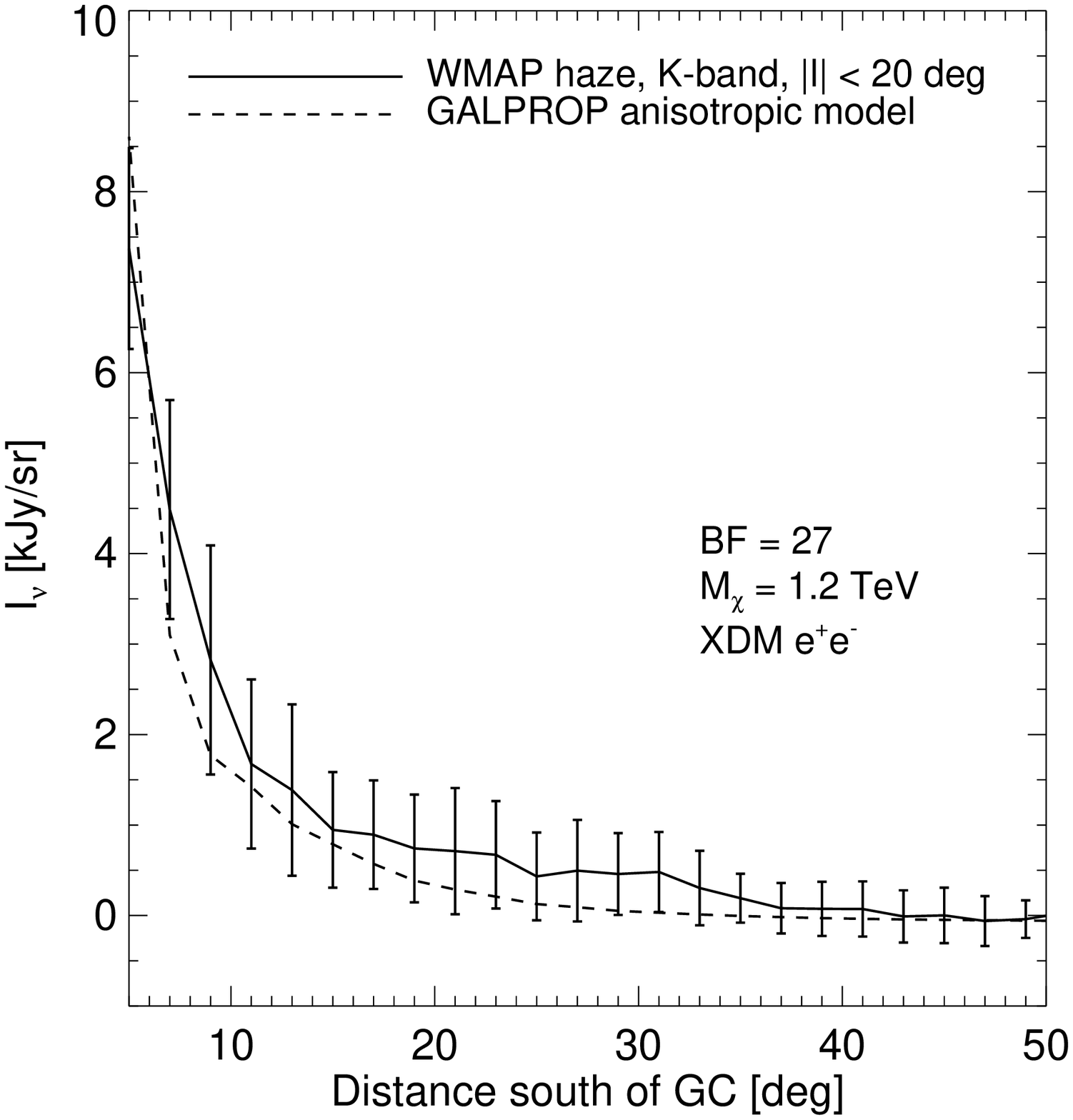}
  }
\caption{
  \emph{Upper left:} A zoom in of the haze from \reffig{compmorph}
  with the two black boxes indicating the area used to plot the
  spectrum shown in the \emph{upper right}.  The light and dark bands
  represent the spectrum of emission in the north and south boxes
  respectively.  The spectrum from DM annihilations (solid lines) is
  quite consistent with the data especially when taking into account
  uncertainties in the IR and starlight components of the ISRF at high
  latitudes.  The dashed and dot-dashed lines are for the same model
  with varying IR and starlight intensities (20\% and 50\% reduction
  in IR and starlight intensities respectively).  The required boost
  factor is nearly identical to that found by fitting the radial
  profile south of the GC to the \emph{Fermi} 4 GeV data (\emph{bottom
    left}) as well independently fitting the microwave haze profile at
  WMAP K-band (\emph{bottom right}).
}
\label{fig:speccomp}
\epm

Lastly, we compare the spectrum of the observed \emph{Fermi} haze to
that produced by the IC emission from $\epp$ generated by the XDM
electrons annihilation channel.  We plot the ${\mathcal H}(E)$
emission in the window defined by $|\ell| < 20$ deg and $10 < |b| <
50$ deg.  This region is dominated by the \emph{Fermi} haze and is
relatively free of other foregrounds.  When comparing the spectra, it
is important to keep in mind that ${\mathcal H}(E)$ has the
\emph{Fermi} $E_{0.5}^{1.0}$ map times $A_{\rm loE}(E)$ removed, and
so in \reffig{speccomp} we show the intensity versus $E$ for
${\mathcal H}(E)$ and $G(E) - A_{\rm loE}(E) \times G_{0.5}^{1.0}$.

Performing an independent fit of the \emph{Fermi} and WMAP haze
profile (intensity as a function of latitude south of the GC) we find
that the required BF for the \emph{Fermi} haze at 4 GeV is BF=24 while
at WMAP 23 GHz it is a nearly identical BF=27, as shown in the bottom
panels of \reffig{speccomp}. In our calculations of the synchrotron
radiation emission, we take into account the presence of both the
ordered and the irregular B-field components.  In the upper right
panel of \reffig{speccomp}, the predicted DM spectrum is plotted over
the \emph{Fermi} data \emph{assuming} a BF of $\sim$24.  It is clear
from the figure that the DM spectrum with this BF provides an
excellent agreement with the data, especially taking into account
uncertainties in the optical and IR ISRF at latitudes far above the
plane. While the cross section in the inner galaxy is roughly a factor
of three lower than that needed to explain local cosmic ray excesses,
this could naturally arise from a radius dependent velocity dispersion
\citep{Cholis:2009va}, or from a depletion of substructure in the
inner galaxy \citep{toroinp}.

%%%%%%%%%%%%%%%%%%%
%%  Conclusions  %%
%%%%%%%%%%%%%%%%%%%

\section{Conclusions}
\label{sec:conclusions}

We have developed a model of Galactic cosmic-ray diffusion that
incorporates both an ordered and turbulent magnetic field component.
The ordered component results in anisotropic diffusion of cosmic-ray
electrons along field lines.  Combining this model of diffusion with
dark matter annihilations in a prolate Galactic dark halo produces an
inverse Compton gamma-ray signal that matches the morphology and
spectrum of the observed \emph{Fermi} gamma-ray haze.  Namely, an
oval-shaped haze with axis ratio $\approx2.0$, extending up to
$|b|\sim50$ deg, and with a cosmic-ray injection spectrum $E^2 dN/dE
\propto E^2$.

The detailed morphology of the haze at low latitudes is still
uncertain.  We have shown that the dust-column to $\pi^0$ gamma-ray
ratio is higher in an ``X'' shaped morphology towards the center of
the Galaxy and that using a map of dust column like the SFD dust map
as a tracer of $\pi^0$ gammas results in an over-subtraction of the
``X''.  The end result is that an oval-shaped haze may then appear
more ``hourglass'' or ``bubble'' shaped.  Using the 0.5-1.0 GeV
\emph{Fermi} map itself (which contains very little of the gamma-ray
haze) as a tracer of disk emission at higher energies is immune to
these line of sight effects and produces a more oval-shaped haze at
the cost of noisier residuals.

Regardless, a three component model of anisotropic diffusion with dark
matter annihilations in a prolate halo plus the \emph{Fermi} 0.5-1.0
GeV map plus a uniform background provides an excellent fit to the
data from 1-20 GeV.  The self-annihilation cross section required for
the dark matter generated IC component is $\sim9\times10^{-25}$
cm$^3$/s (boost factor $\sim30$), which is easily obtainable via the
Sommerfeld enhancement in our models and also produces the microwave
haze.  Furthermore, this boost factor is well within the bounds of
thermal relic and CMB constraints \citep{Slatyer:2009yq,
  Zavala:2009mi}.

The most significant outstanding issues are the sharp ``edges'' of the
haze at high latitudes and also the morphology of the haze at low
latitudes.  Sharp edges are not particularly expected with either a
dark matter annihilation or astrophysical (such as winds or jets)
mechanism, unless the spectrum at the edge is significantly hardened
as does not appear to be the case.  Magnetic confinement could
potentially help both explanations, though care must be taken not to
significantly synchrotron brighten the edges which are not seen in the
WMAP microwave data.  The low latitude morphology of the haze
(``oval'' versus ``bubble'' shape) may become more clear as more data
are collected by \emph{Fermi}.  In particular, at high energies, the
disk fades much more quickly than the haze because of the softer
spectrum of the disk, and so the low latitude haze may be revealed at
high energies with 5 to 10 more years of data.

Of course, there is the possibility of a hybrid scenario in which some
event evacuates a cavity towards the Galactic center that is filled
with high energy electrons from dark matter annihilation that are
trapped by magnetic confinement.  \cite{Su:2010qj} discount this
possibility under the assumption that the dark matter signal would be
more spherical, but we have shown here that this is not the case in
general for triaxial halos.  Injection from dark matter annihilation
would also have the advantage that the hard spectrum can be obtained
(as we have shown in this paper) and the injection is \emph{extended}.
Nevertheless, inside the edge, the haze appears to have a profile that
is roughly flat in latitude above $|b|>30^o$.  Such a projected
profile seems nearly impossible to realize (either with astrophysical
or dark matter models) unless electrons pile up on the edges, though
the naive expectation would be that the gammas would be
limb-brightened which is not observed.  If these features persist in
future data, such hybrid scenarios are inevitable.

Lastly, we point out that introducing a significant ordered field as
we have could potentially produce a significantly polarized microwave
signal.  By design, our model does reproduce the observed microwave
haze in total temperature; comparison with the WMAP polarization data
will be the subject of future work.

%%%%%%%%%%%%%%%%%%%%%%%
%%  Acknowledgments  %%
%%%%%%%%%%%%%%%%%%%%%%%

\section{Acknowledgments}

The authors would like to thank Douglas Finkbeiner, Joseph Gelfand,
and Ronnie Janson for helpful conversations.  We especially thank Lisa
Goodenough for providing valuable insights with respect to anisotropic
propagation using GALPROP.  GD is supported by the Harvey L.\ Karp
Discovery Award. IC has been partially supported by DOE OJI grant \#
DE-FG02-06ER41417, and also by the Mark Leslie Graduate
Assistantship. NW is supported by DOE OJI grant \#DE-FG02-06ER41417
and NSF grant \#0947827, as well as by the Amborse Monell Foundation.

%%%%%%%%%%%%%%%%%%%%
%%  Bibliography  %%
%%%%%%%%%%%%%%%%%%%%

%%%%%%%%%%%%%%%%%%
%%  Appendix A  %%
%%%%%%%%%%%%%%%%%%

\appendix
\section{Anisotropic diffusion in the GALPROP code}
\label{app:modgalprop}

In GALPROP, the diffusion equation is solved through the
Crank-Nicholson implicit method \citep{Strong:2007nh,
  1992nrfa.book.....P}:
\beq
  \frac{\partial \psi_{i}}{\partial t} = \frac{\psi_{i}^{t+\Delta t} -
    \psi_{i}^{t}}{\Delta t} = \frac{\alpha_{1}\psi_{i-1}^{t+\Delta t}
    - \alpha_{2}\psi_{i}^{t+\Delta t} + \alpha_{3}\psi_{i+1}^{t+\Delta
      t}}{\Delta t} + Q_{i},
\label{eq:Crank_Nicholson}
\eeq
where $i$ is the index of position ($r$ or $z$) or momentum and
$\alpha_{1,2,3}/\Delta t$ are the Crank-Nicholson coefficients.  In
the case where $D$ is homogeneous in space, these coefficients are:
\beq
  \frac{\alpha_{1}}{\Delta t} = D \frac{2r_{i} - \Delta
    r}{2r_{i}(\Delta r)^{2}}, \ \frac{\alpha_{2}}{\Delta t} = D
  \frac{2r_{i}}{r_{i}(\Delta r)^{2}}, \ \frac{\alpha_{3}}{\Delta t} =
  D \frac{2r_{i} + \Delta r}{2r_{i}(\Delta r)^{2}},
\eeq 
for diffusion along $r$ and
\beq
 \frac{\alpha_{1}}{\Delta t} = \frac{D}{(\Delta z)^{2}}, \
 \frac{\alpha_{2}}{\Delta t} = \frac{2D}{(\Delta z)^{2}}, \
 \frac{\alpha_{3}}{\Delta t} = \frac{D}{(\Delta z)^{2}}
\eeq
for diffusion along $z$.  With the new terms from anisotropic
diffusion (see Eq.\ \ref{eq:Diffusion_termGeneral}) the
Crank-Nicholson coefficients become:
\beq
  \frac{\alpha_{1}}{\Delta t} = D_{rr_{i}} \frac{2r_{i} - \Delta r}
       {2r_{i}(\Delta r)^{2}} -
       \frac{D_{rr_{i+1}}-D_{rr_{i-1}}}{4(\Delta r)^{2}}, \ 
  \frac{\alpha_{2}}{\Delta t} = D_{rr_{i}}
  \frac{2r_{i}}{r_{i}(\Delta r)^{2}}, \
  \frac{\alpha_{3}}{\Delta t} = D_{rr_{i}} \frac{2r_{i} + \Delta
    r}{2r_{i}(\Delta r)^{2}} +
  \frac{D_{rr_{i+1}}-D_{rr_{i-1}}}{4(\Delta r)^{2}}, 
\label{eq:Crank_Nicholson_coefficients1}
\eeq
for diffusion along $r$ and
\beq
  \frac{\alpha_{1}}{\Delta t} = \frac{D_{zz_{i}}}{(\Delta z)^{2}} -
  \frac{D_{zz_{i+1}}-D_{zz_{i-1}}}{4(\Delta z)^{2}}, \
  \frac{\alpha_{2}}{\Delta t} = \frac{2D_{zz_{i}}}{(\Delta z)^{2}}, \
  \frac{\alpha_{3}}{\Delta t} = \frac{D_{zz_{i}}}{(\Delta z)^{2}} +
  \frac{D_{zz_{i+1}}-D_{zz_{i-1}}}{4(\Delta z)^{2}},
\label{eq:Crank_Nicholson_coefficients2}
\eeq
for diffusion along $z$, and where, as in the main text, we have taken
$D_{rz} = D_{zr} = 0$ (since $B_{r} = 0$) for simplicity.  When
iterating recursively for a steady-state $\psi$, GALPROP utilizes the
fact that the $r$ and $z$ directions are separable, whereas if we take
$D_{rz} = D_{zr} \ne 0$, these directions are not separable.  So while
we use Eqs.\ \ref{eq:Crank_Nicholson_coefficients1} and
\ref{eq:Crank_Nicholson_coefficients2} in practice, the general
quantization of Eq.\ \ref{eq:Diffusion_termGeneral} is (superscripts
are spatial indices for clarity)
\begin{eqnarray}
  \overrightarrow{\nabla}(D\overrightarrow{\nabla} \psi) &=&
  \left(\frac{D_{rr}^{i,j}}{r^{i}} + \frac{D_{rr}^{i+1,j} -
    D_{rr}^{i-1,j}}{2\Delta r} + \frac{D_{rz}^{i,j+1} -
    D_{rz}^{i,j-1}}{2\Delta z}\right) \times \frac{\psi^{i+1,j} -
    \psi^{i-1,j}}{2\Delta r} + \nonumber \\ & &
  \left(\frac{D_{rz}^{i,j}}{r^{i}} + \frac{D_{rz}^{i+1,j} -
    D_{rz}^{i-1,j}}{2\Delta r} + \frac{D_{zz}^{i,j+1} -
    D_{zz}^{i,j-1}}{2\Delta z}\right) \times \frac{\psi^{i,j+1} -
    \psi^{i,j-1}}{2\Delta z} + \\ & & D_{rr}^{i,j} \times
  \frac{\psi^{i+1,j} + \psi^{i-1,j} - 2\psi^{i,j}} {(\Delta r)^2} +
  D_{zz}^{i,j} \times \frac{\psi^{i,j+1} + \psi^{i,j-1} - 2\psi^{i,j}}
  {(\Delta z)^2} + \nonumber \\ & & D_{rz}^{i,j} \times
  \frac{\psi^{i+1,j+1} + \psi^{i-1,j-1} -\psi^{i+1,j-1}
    -\psi^{i-1,j+1}} {2\Delta r\Delta z}. \nonumber
\end{eqnarray}

%%%%%%%%%%%%%%%%%%
%%  Appendix B  %%
%%%%%%%%%%%%%%%%%%

\section{Diffusion dependence on magnetic field}
\label{app:diff_dep_B}

Let us consider the generic case of an electron traveling in a
magnetic field with both an irregular $B_{\rm irr}$ and ordered
$B_{\rm ord}$ component.  As the electron spirals around the ordered
field lines with cyclotron frequency $\Omega$, there is a
characteristic frequency $\nu$ at which the electron is scattered from
its path by the irregular component.  In the case of a strong ordered
component $B_{\rm ord} \gg B_{\rm irr}$, $\Omega \gg \nu$ while for
$B_{\rm ord} \ll B_{\rm irr}$, $\Omega \ll \nu$.  In other words, for
strong ordered fields, the electron spirals around the field line many
times before it is deflected by the irregular component.  As noted in
\refsec{anisodiff}, this behavior is written in the diffusion tensor
as \citep{1965P&SS...13....9P}: 
\beq 
  D_{ij} = D_0\left(\frac{\nu^2\delta_{ij} + \Omega_i\Omega_j}{\nu^2 +
    \Omega^2}\right), \eeq which, for the case of an ordered field
  completely along the $z$-direction, leads to the relation \beq
  \frac{D_{rr}}{D_{zz}} \propto \frac{1}{1+B_{\rm ord}^2/B_{\rm
      irr}^2}.
  \label{eq:drrdzz}
\eeq

We wish to motivate Equation \ref{eq:drrdzz} from the perspective of
diffusion \emph{lengths} $\lambda$ since that it the most direct
measure of the diffusion coefficient, $D_{ij} = \lambda_{ij}c/3$
(where $c$ is the speed of light).  In our scenario then,
\begin{equation}
  \frac{\lambda_{zz}}{\lambda_{rr}} \sim \frac{r_{gyr}N}{r_{gyr}}
  \sim N,
\label{eq:lamba_1}
\end{equation}
where $r_{gyr}$ is the gyroradius and $N$ is the number of scatterings
of the particle by angle $\phi \sim B_{\rm irr}/B_{\rm tot}$, which is
the inclination angle of the field lines from the direction of the
mean field due to irregularities \citep[see][]{Longair}.  Note that,
in the case of $B_{\rm irr} \gg B_{\rm ord}$, $\phi$ is large and
particles are deflected significantly from their initial direction
within one gyroradius while for $B_{\rm irr} \ll B_{\rm ord}$, $\phi
\ll 1$ and particles follow the field lines of the local ordered
field.  For the particle to scatter by $\sim$1 radian, we need
$\sqrt{N}\phi\sim1$ which implies
\beq
  \frac{\lambda_{zz}}{\lambda_{rr}} \sim \phi^{-2} \sim
  B_{\rm tot}^{2}/B_{\rm irr}^{2},
\eeq
and since $B_{\rm tot} = \sqrt{B_{\rm ord}^2 + B_{\rm irr}^2}$, 
\beq
  \frac{\lambda_{zz}}{\lambda_{rr}} \sim \phi^{-2} \sim (1 +
  B_{\rm ord}^{2}/B_{\rm irr}^{2})
\eeq
as desired.

%%%%%%%%%%%%%%%%%%
%%  Appendix C  %%
%%%%%%%%%%%%%%%%%%

\section{Low frequency radio emission}
\label{app:low_freq}

\bpm
  \centerline{
    \includegraphics[width=0.49\textwidth]{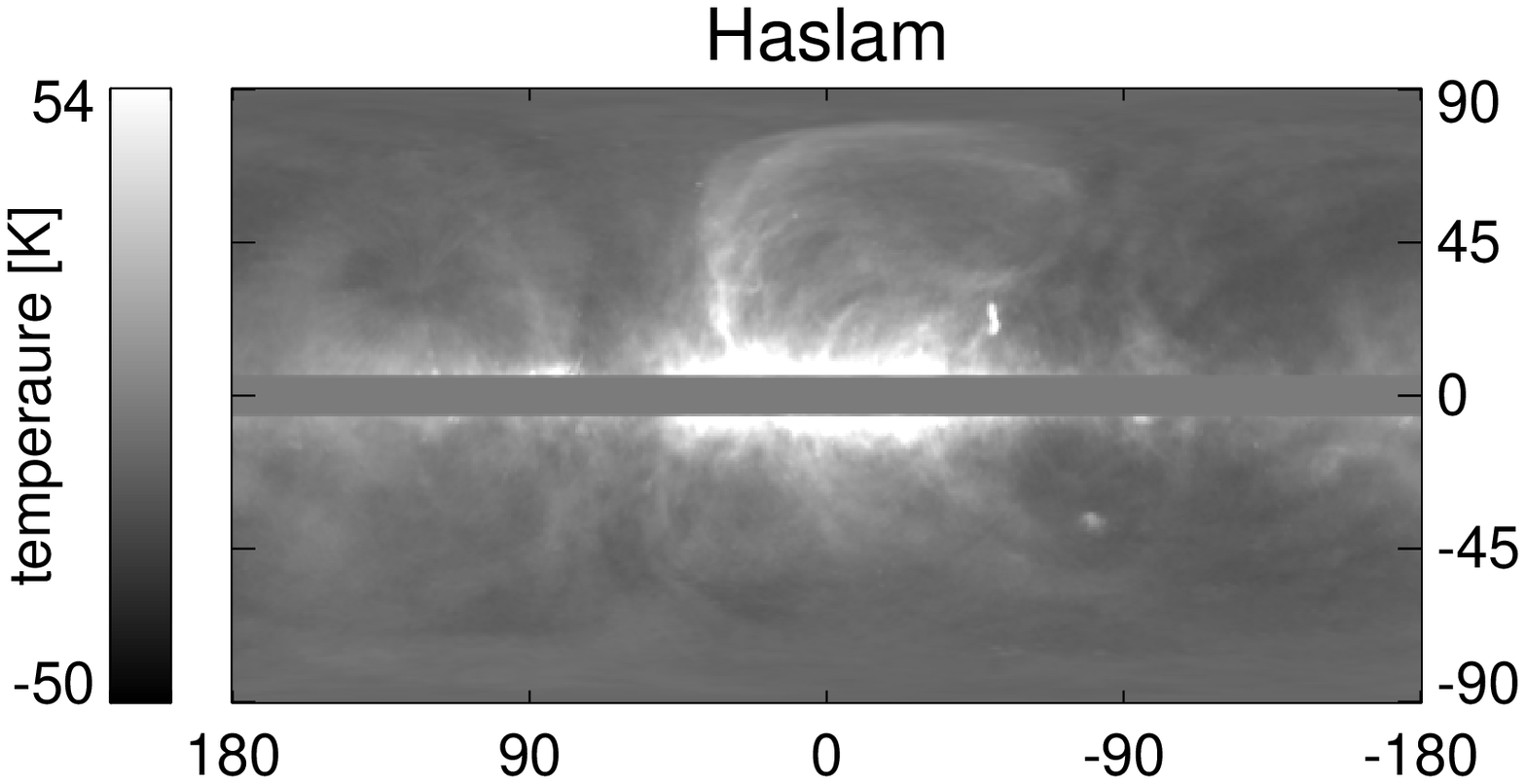}
    \includegraphics[width=0.49\textwidth]{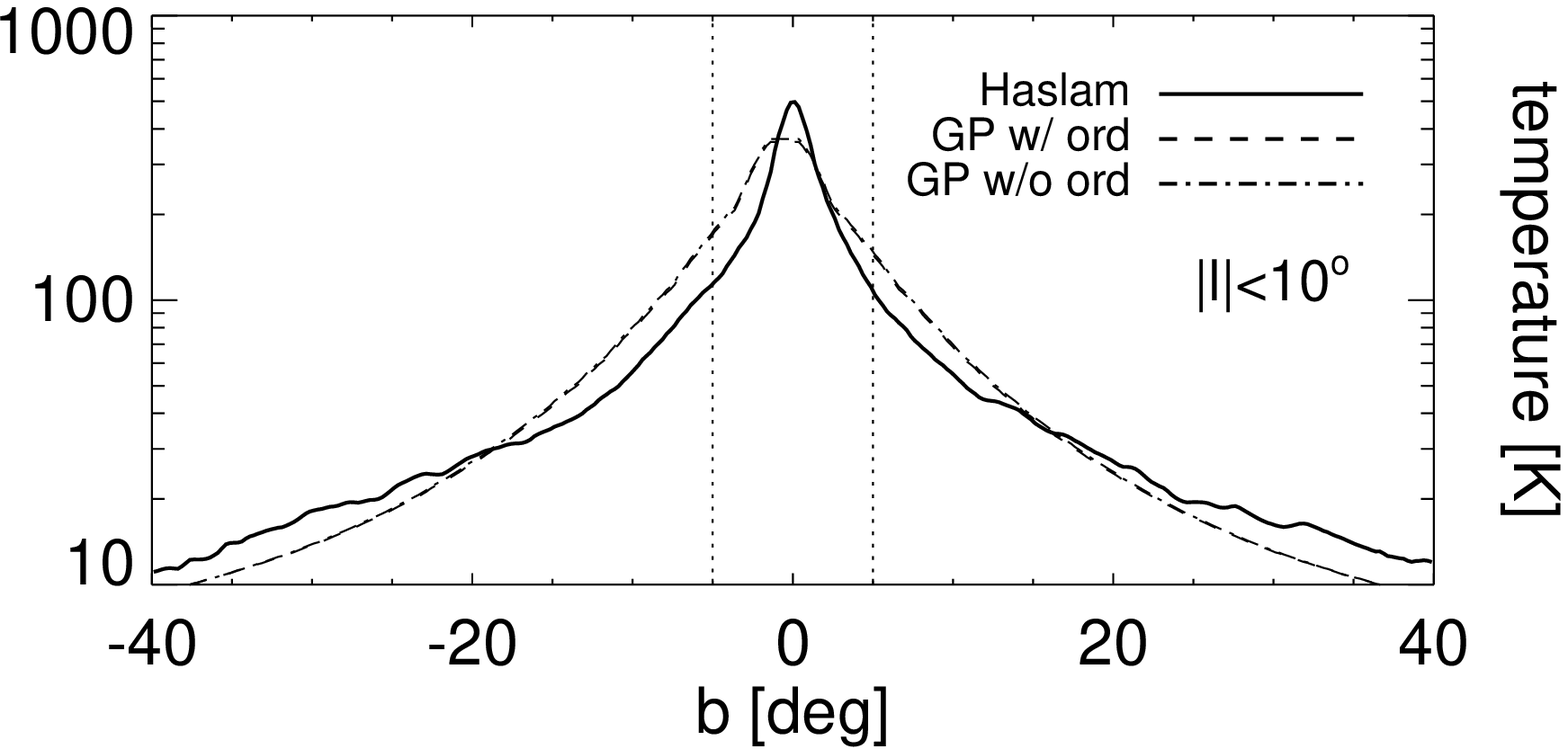}
  }
  \centerline{
    \includegraphics[width=0.49\textwidth]{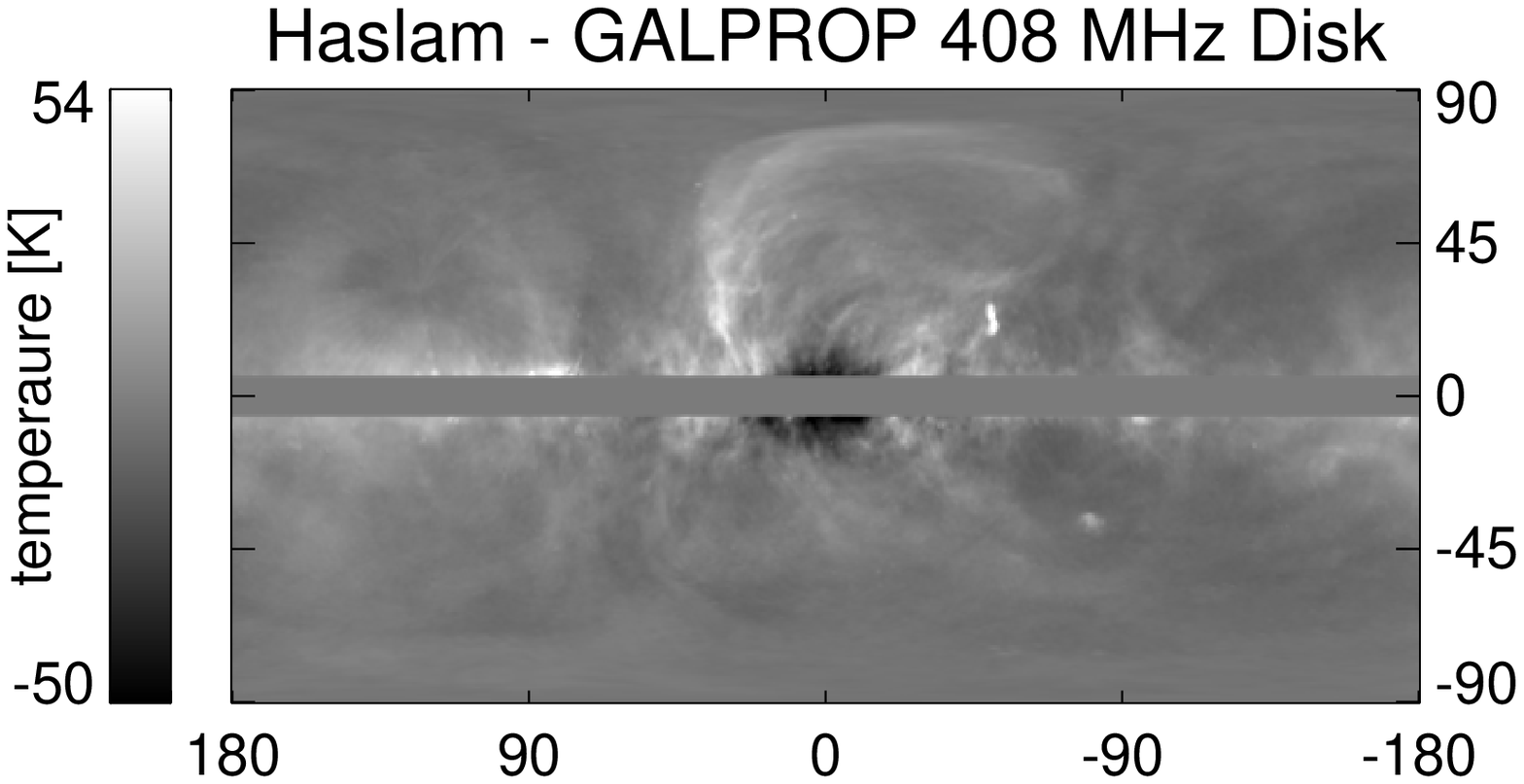}
    \includegraphics[width=0.49\textwidth]{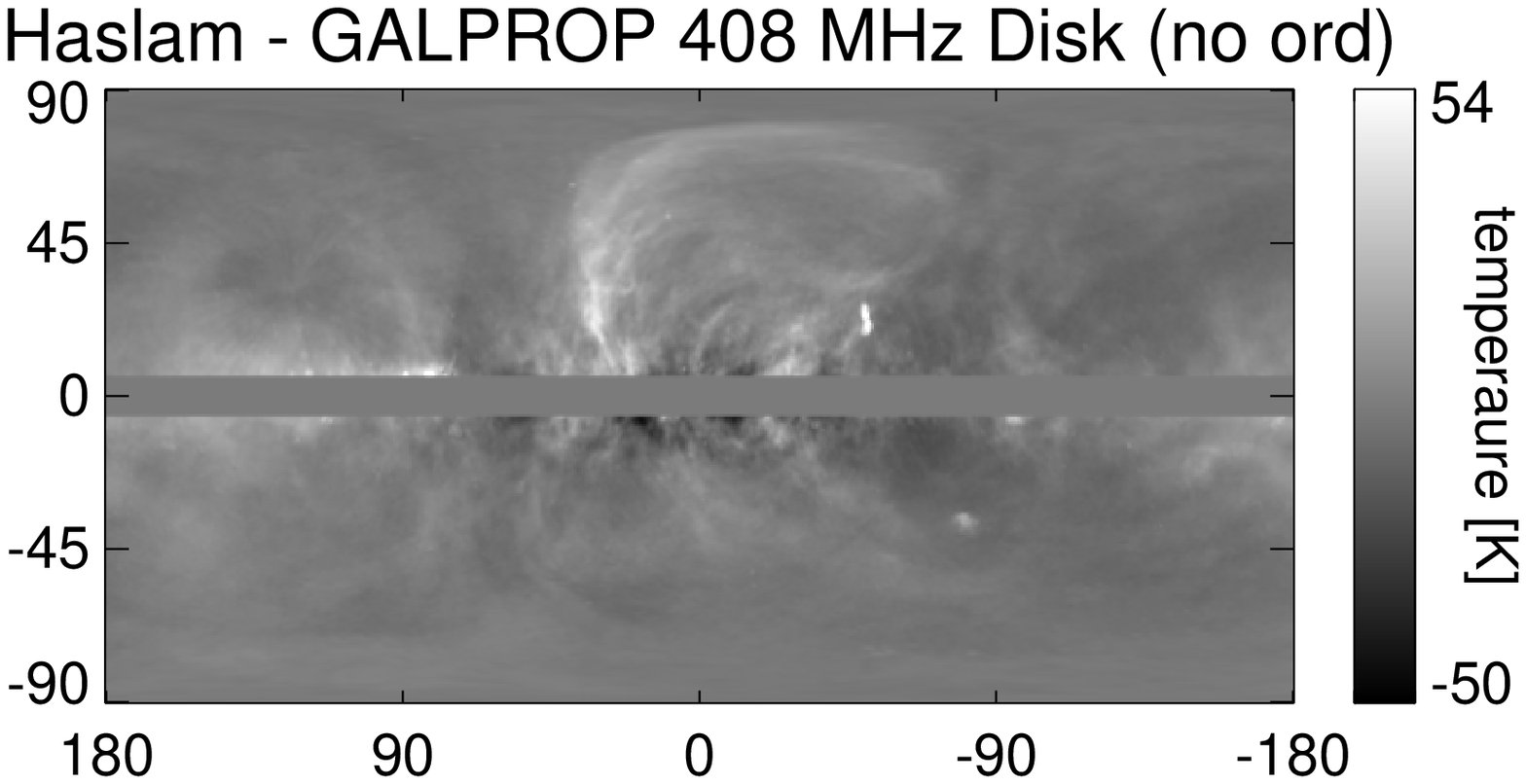}
  }
  \centerline{
    \includegraphics[width=0.49\textwidth]{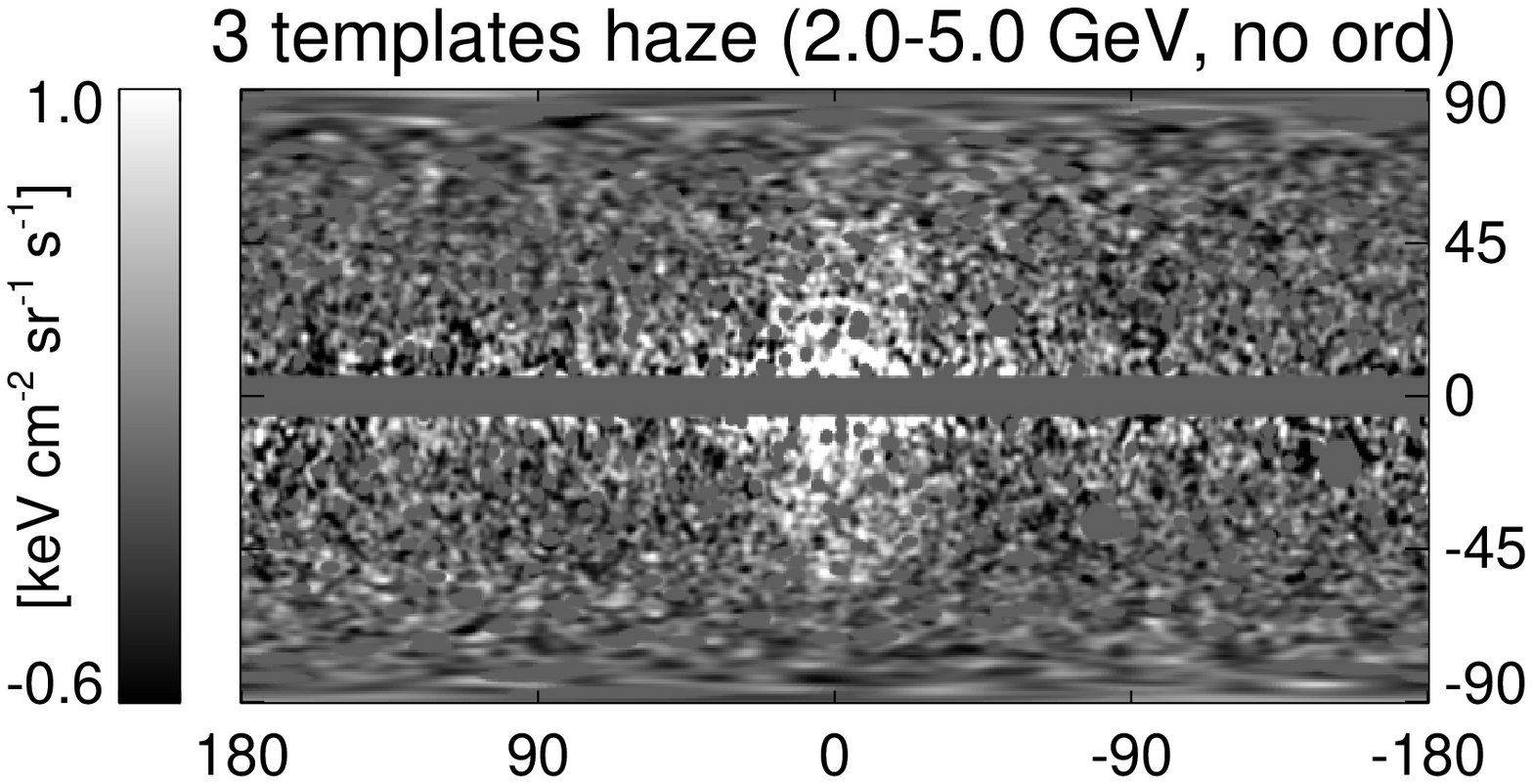}
    \includegraphics[width=0.49\textwidth]{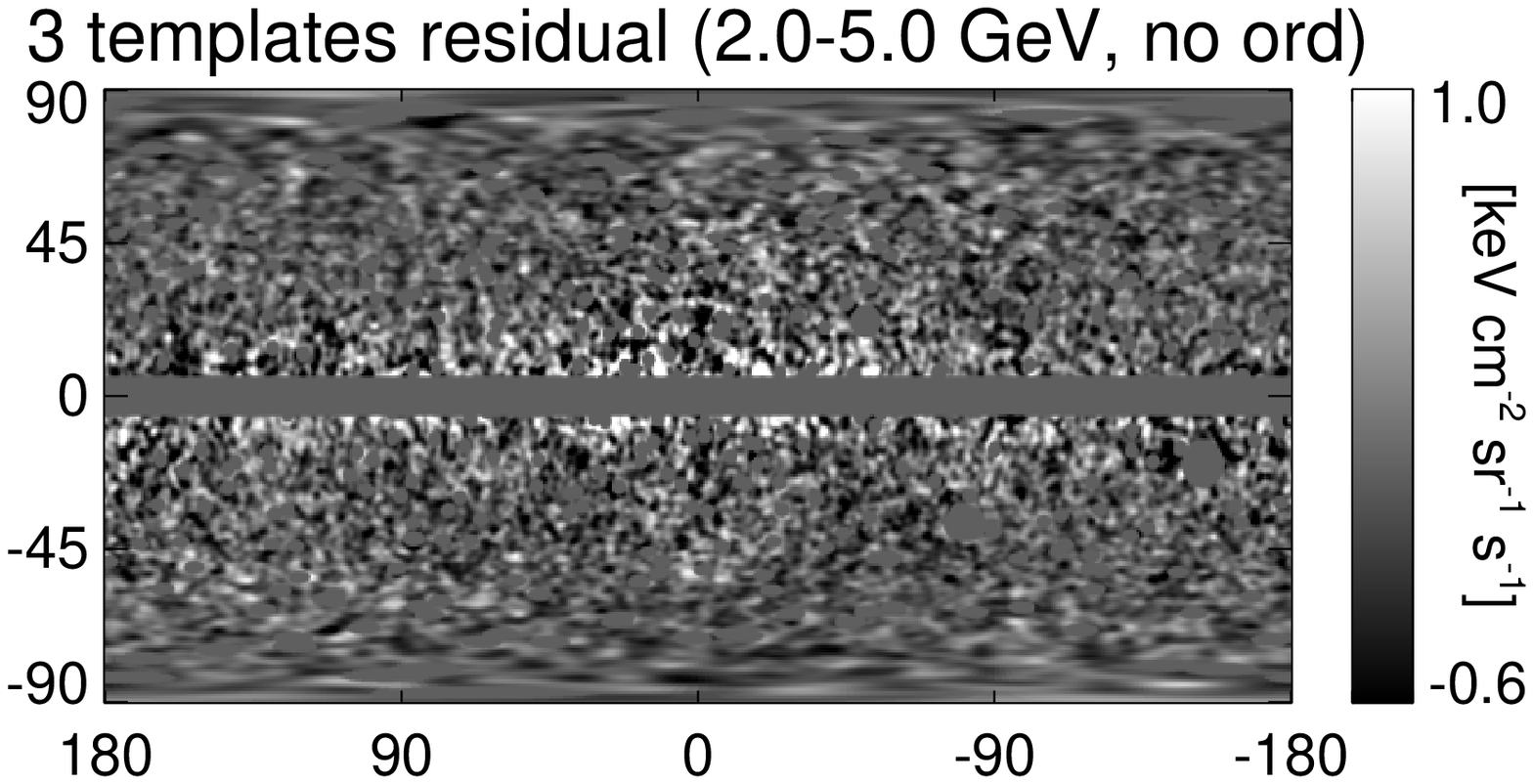}
  }
\caption{
  \emph{Upper left:} The full sky Haslam 408 MHz map.  \emph{Upper
    right:} the latitudinal profile of Haslam and our 408 MHz
  synchrotron GALPROP map for primary electrons injected via SN shocks
  using our anisotropic diffusion model (both with and without the
  ordered field component).  \emph{Middle left:} Haslam minus our 408
  MHz GALPROP map.  While the agreement is in general quite good, the
  map is slightly too concentrated towards the center.  This is
  alleviated by removing the ordered field (\emph{middle right})
  indicating that more complex ordered field morphologies than our
  simple parameterization allows can provide a good fit to the data.
  In addition to being a good fit to Haslam, the model without an
  ordered field is a reasonable match to the \emph{Fermi} haze, though
  it is a bit centrally concentrated (\emph{bottom panels}).
}
\label{fig:fithaslam}
\epm

Dominated by synchrotron emission from electrons with energies
$\sim$few GeV, the Haslam 408 MHz map \citep{1982A&AS...47....1H}
provides an excellent constraint on both the injection morphology of
the primary electrons via SN shock acceleration and also the magnetic
field morphology.  Thus, it is important to check that our anisotropic
diffusion model maintains the ``disk-like'' shape of synchrotron from
primary electrons at 408 MHz.  \reffig{fithaslam} shows the Haslam
map, and the Haslam map and the predicted 408 MHz synchrotron emission
due to SN injection using our full anisotropic model.  As in
\cite{2010PhRvD..82b3518L}, we normalize the model to Haslam by
setting the total emission in the region $|\ell| \leq 10^o$ and $-90^o
\leq b \leq -5^o$ equal.

As shown in the radial profile panel (upper right), the agreement is
very good, though the map difference (middle left panel) indicates
that the stronger magnetic field in the center (due to the ordered
component) may make the emission somewhat steeper in the region
$\sim$1-2 kpc.  However, we point out that, not only is the gamma-ray
signal dominated by emission at higher latitudes, but we have used a
very simple parameterization for the ordered field and additional
field parameters (to lower the field within the inner 2 kpc) can
remove the discrepancy.  In fact, we can remove the ordered component
altogether and the model comes into close agreement with both Haslam
(middle right panel) and the \emph{Fermi} data (bottom panels; also
cf.\ \reffig{addanisotropy}).

\end{document}